\documentclass[reprint,amsmath,amssymb,aps,superscriptaddress,twocolumn,10pt]{revtex4-1}
\usepackage[colorlinks=true,allcolors=blue]{hyperref}
\usepackage{graphicx,color}
\usepackage{dcolumn}
\usepackage{subfigure}
\usepackage{bm}
\usepackage{amsmath,amsfonts,amssymb}
\usepackage{acronym}
\usepackage{enumitem}
\usepackage{array}
\usepackage{multirow}
\usepackage{CJK}
\usepackage{ulem}
\usepackage[thicklines]{cancel}
\allowdisplaybreaks
\newcommand{\be}{\begin{equation}}
\newcommand{\ee}{\end{equation}}
\newcommand{\bea}{\begin{eqnarray}}
\newcommand{\eea}{\end{eqnarray}}
\newcommand{\nn}{\nonumber}

\newcommand{\TRC}{MOE Key Laboratory of TianQin Mission, TianQin Research Center for Gravitational Physics $\&$ School of Physics and Astronomy, Frontiers Science Center for TianQin, CNSA Research Center for Gravitational Waves, Sun Yat-sen University (Zhuhai Campus), Zhuhai 519082, China}
\newacro{GW}{gravitational wave}
\newacro{GR}{general relativity}
\newacro{EMRI}{Extreme Mass-Ratio Inspiral}
\newacro{EdGB}{Einstein-dilaton-Gauss-Bonnet}
\newacro{GB}{Gauss-Bonnet}
\newacro{BH}{ black hole}
\newacro{MBH}{massive black hole}
\newacro{MBHB}{Massive Black Hole Binary}
\newacro{SCO}{stellar mass compact object}
\newacro{NK}{numerical kludge}
\newacro{PE}{parameter estimation}
\newacro{SNR}{signal-to-noise ratio}
\newacro{PN}{post newtonion}
\newacro{FIM}{Fisher Information Matrix}
\newacro{LVK}{LIGO, Virgo and KAGRA}

\begin{document}

\title{Constraining the EdGB Theory with Extreme Mass-Ratio Inspirals}

\author{Jing Tan}
\author{Jian-dong Zhang}
\email{zhangjd9@mail.sysu.edu.cn}
\author{Hui-Min Fan}
\author{Jianwei Mei}
\affiliation{\TRC}

\begin{abstract}
The \ac{EdGB} theory is a modified theory of gravity which include a scalar field to couple with the higher order curvature terms.
It has already been constrained with various observations include the \ac{GW} with \ac{LVK} Collaboration.
In this work, we study the capability for space-borne \ac{GW} detectors to constrain the \ac{EdGB} theory using the signal of \acp{EMRI}.
We use the ``\ac{NK}'' method to construct the waveform of \ac{EMRI} in the \ac{EdGB} theory,
focusing on the case when the central black hole is spinless.
We then study how a future space-borne gravitational wave detector, TianQin, for example,
can place constraints on the \ac{EdGB} theory through the detection of \acp{EMRI}.
With the analysis using mismatch and \ac{FIM},
we find that the \ac{EdGB} parameter $\sqrt{\alpha}$ is expected to be constrained to the level of $\sim\mathcal{O}(0.1)$ km.
\end{abstract}

\maketitle

\section{Introduction}\label{sec:intro}

The 2030s are expected to become an exciting period when the space-borne \ac{GW} detectors,
including TianQin \cite{Luo:2015ght,TianQin:2020hid}, LISA \cite{Danzmann:1997hm, LISA:2017pwj}, and Taiji\cite{Hu:2017mde}, will be put into operation.
These detectors will for the first time open the unexplored millihertz frequency band of \ac{GW} spectrum,
observing a lot of new \ac{GW} sources, such as Galactic Compact Binaries \cite{Korol:2017qcx, Huang:2020rjf},
\acp{MBHB} \cite{Klein:2015hvg,Wang:2019ryf,Feng:2019wgq},
the inspiral of Stellar Mass Black Hole Binaries \cite{Sesana:2016ljz, Kyutoku:2016ppx,Liu:2020eko}, \acp{EMRI} \cite{Babak:2017tow,Fan:2020zhy},
and Stochastic \ac{GW} Background \cite{Caprini:2015zlo, Bartolo:2016ami, Liang:2021bde, Cheng:2022vct},
and are promising significant boost to fundamental physics \cite{Gair:2012nm, Barausse:2020rsu, LISA:2022kgy}, astrophysics \cite{LISA:2022yao, Baker:2019pnp} and cosmology \cite{Tamanini:2016zlh, Zhu:2021aat,LISACosmologyWorkingGroup:2022jok, Caldwell:2019vru}.
One of the most important topics is to study the nature of gravity and \acp{BH} in the strong field regime \cite{Shi:2019hqa,Shi:2022qno,Zi:2021pdp,Xie:2022wkx,Rahman:2022fay}.
With future space-borne detectors, not only the precision of some existing tests can be significantly improved,
but also some currently inaccessible types of tests become possible.
To fully employ the opportunity, it is necessary to further explore and prepare for the tests of \ac{GR} that can be done with future space-based detectors.

\ac{EMRI} is one of the most important targets for a space-borne \ac{GW} detector.
It consisted of a \ac{SCO} orbiting around a \ac{MBH} in the strong field region.
In an \ac{EMRI} event, the \ac{SCO} can orbit the \ac{MBH} for $\mathcal{O}(10^5)$ cycles before the plunge
and can map the surrounding geometry of the \ac{MBH} to unparalleled precision \cite{Barack:2003fp,Amaro-Seoane:2007osp,Zi:2021pdp,AbhishekChowdhuri:2023wdf}.
This feature can be used to test many aspects of gravity and \acp{BH},
such as telling the difference between \ac{GR} and alternative theories of gravity ,
or testing the no-hair theoreom of \ac{BH}.
In this work we will focus on the \ac{EdGB} theory \cite{Kanti:1995vq, Torii:1996yi, Kanti:1997br, Nojiri:2010wj, Pomazanov:2003wq, Pani:2009wy},
which is a modified theory of gravity including a dilation coupled to the Gauss-Bonnet term, to make it non-trivial in the $4D$ case.
The coupling constant $\alpha$ has the dimension of length squared, and if it equals zero, this theory will reduce to \ac{GR}.

With the motion of the Cassini spacecraft, the upbound of $\sqrt{\alpha}$ is constrained to the level of $10^7$km within the PPN framework\cite{Amendola:2007ni}.
By observing the orbital decay rate of the \ac{BH} low mass X-ray binaries, $\sqrt{\alpha}$ can be constrained to $1.9$km \cite{Yagi:2012gp}.
The detection of \ac{GW} provide tighter constraints \cite{Nair:2019iur, Yamada:2019zrb, Tahura:2019dgr, Carson:2020ter, Okounkova:2020rqw, Perkins:2021mhb, Wang:2021jfc, Wang:2023wgv, Odintsov:2020xji} to the level of $\mathcal{O}(1)$km.
With future space-borne \ac{GW} detectors, the inspiral \cite{Shi:2022qno,Luo:2024vls} and merger \cite{Shao:2023yjx} of \ac{MBHB} can also put the constraint on $\sqrt{\alpha}$ to the level of $10^4$km.
An inspiral-merger-ringdown (IMR) consistency tests is also achieved in \cite{Carson:2020cqb} for stellar mass binaries with multiband observations.
Most of the analysis are achieved within the parameterized post Einstein framework, and the correction is given by the post-Newtonian method.
Thus it's only suitable for the cases with comparable masses, and can not be used for \acp{EMRI}.
However, a direct extension shows that \ac{EMRI} can give a better constraint to the level of $\mathcal{O}(0.1)$km.

In this work, we focus on using \ac{EMRI} to constrain the \ac{EdGB} theory.
Based on the static \ac{BH} solution in this theory, we develop the waveform model using the \ac{NK} method.
In the generation of the waveform, three factors need to be taken into consideration:
(1) the change of the metric, and thus the modification of the geodesic equation;
(2) the change of the energy flux and angular momentum flux of the \ac{GW}, and thus the evolution of the orbit;
(3) the change of the field equation, and thus the generation of the waveform.
Since the metric of \ac{BH} is spherical symmetric, we will only consider the equatorial eccentric case in the waveform generation.

Then, we study the capability of TianQin to place constraints on the EdGB theory through the detection of EMRI signals.
This is a space-based \ac{GW} detector expected to be launched around 2035 \cite{Luo:2015ght, TianQin:2020hid}.
By calculating the mismatches between the waveform in \ac{GR} and \ac{EdGB}, we find that it will beyond the threshold for $\sqrt{\alpha}$ at the level of 0.1km.
The analysis with \ac{FIM} can also give the similar constraint.

The paper is organized as follows.
In section \ref{sec:waveform}, we describe the construction of the \ac{EMRI} waveforms in the \ac{EdGB} theory assuming that the central \ac{MBH} is spinless.
In section \ref{sec:method}, we introduce the response and noise for the detector, and the method to estimate the constraint and compare the waveform.
In section \ref{sec:result}, we present the result on the projected constraint on the \ac{EdGB} theory that can be achieved with TianQin.
We give a summarization in \ref{sec:summary}.
Throughout this paper, we assume $G=c=1$.

\section{EMRI waveforms in the EdGB theory} \label{sec:waveform}

In this section, we describe the construction of the \ac{EMRI} waveforms in the \ac{EdGB} theory.
In the following subsections, we will review the basic ideas of the \ac{NK} method,
briefly introduce the \ac{EdGB} theory, and specify the metric of a spinless black hole that we will use for the central \ac{MBH}.
Then we will describe the geodesic equations for test particles moving on the \ac{MBH} background,
and describe the generation of the \ac{EMRI} waveforms.

The calculation of the waveform for a general \ac{EMRI} system is challenging.
In the GR case, various waveform models have been established.
The gravitational self-force (GSF) computation which employs black hole perturbation theory serves as an ideal tool for the development of high-accuracy GWs from EMRI \cite{Poisson:2011nh, Osburn:2015duj, Warburton:2017sxk, vandeMeent:2017bcc}.
However, to calculate the self-force at the post-adiabatic order for general \ac{EMRI} system is still a big question.
To meet the data analysis needs of EMRI detection, it is necessary to generate a large number of waveforms.
This requires fast waveform calculations, giving rise to approximate Kludge waveforms that contain analytic kludge (AK) \cite{Barack:2003fp}, \ac{NK} \cite{Babak:2006uv}, and augmented analytic kludge (AAK) waveforms \cite{Chua:2017ujo,Liu:2020ghq}.
Recently, the Fast MERI Waveform (FEW) \cite{Katz:2021yft} is also developed to generate the waveform faster.
AK model aims to capture the main feature of \acp{EMRI}, considering particle moving on a Keplerian orbit.
Due to the inaccuracy on the evolution of orbital frequencies,
the AK model can not fits relativistic waveform models very well.
But it's still widely used in the mock data and data analysis studies,
since it can be generated very fast for general systems.
The more accurate NK model considers particle moving on the geodesic with the orbit evolution in the PN framework.
The AAK method using the information from NK model to calibrate the evolution of the orbit in AK model,
and thus the waveform can be generated fast and accurate.
The AAK waveform maintains a lower computational cost while providing a relatively accurate waveform, but this is based on the NK waveform.
The NK method is used in our study.

The construction of the NK model is typically decomposed into several components. Firstly, one can construct constants array $(E, L_z, Q)$ for a particle on a certain orbit, representing energy, axial angular momentum, and Carter constant respectively. The flux corresponding to these constants will lead to their evolution, and this process will continue repetitively to produce a inspiral trajectory. The flux can be written as \cite{Gair:2005ih}:
\begin{eqnarray}
\dot{E} &=& \mu f_E(p,e,\iota), \\
\dot{L_z} &=& \mu f_{L_z}(p,e,\iota), \\
\dot{Q} &=& \mu f_Q(p,e,\iota).
\end{eqnarray}
Where $\mu$ is the mass of the secondary, the parameters of the orbit $(p,e,\iota)$ are the semi-latus rectum, eccentricity, and inclination angle respectively.
The orbit without radiation reaction is given by the geodesic equations:
\be
\frac{ dr}{ d\lambda} = \pm\sqrt{V_r},~
\frac{ d\theta}{ d\lambda} = \pm\sqrt{V_\theta},~
\frac{ d\phi}{d\lambda} = V_{\phi},~
\frac{ dt}{ d\lambda} = V_t.
\ee
The parameters of the orbit can be obtained from the coordinates using Keplerian definitions.
Then by integrating these equations together, we can get the inspiral trajectory in Boyer-Lindquist coordinates.
However, since the time scale for the evolution of the orbit is much larger than the orbit period, in practical, the secondary will move on one geodesic for several periods, then we evolve the moving constants, and it will move on another geodesic.
More technical details can be found in \cite{Babak:2006uv}.
Finally, by mapping the Boyer-Lindquist coordinates $(r, \theta, \phi,\tau=\lambda/\mu)$ to spherical polar coordinates in a flat-space setting, one can construct the waveform in quadrupole approximation.

\subsection{The EdGB theory and its black hole solution}

\ac{EdGB} is a special example of quantum gravity-inspired theories featuring quadratic curvature terms in the action \cite{Kanti:1995vq, Torii:1996yi, Kanti:1997br, Pomazanov:2003wq, Pani:2009wy},
\be
\mathcal{S}=\int d^4x\sqrt{-g}\left[\frac{R}{16\pi} -\frac{1}{2}\partial_\mu\vartheta\partial^\mu\vartheta +\alpha\frac{e^{\vartheta}}{4}\mathcal{R}^2_{GB}+\mathcal{L}_m\right],
\ee
$\alpha$ is the coupling constant between the dilaton field $\vartheta$ and the \ac{GB} term $\mathcal{R}^2_{GB}$,
which is
\be
\mathcal{R}^2_{GB}=(R^2-4R_{\mu\nu}R^{\mu\nu}+R_{\mu\nu\sigma\rho}R^{\mu\nu\sigma\rho}).
\ee
The equations of motion for the \ac{EdGB} theory is
\bea
G_{\mu\nu} + 16\pi\alpha
{\cal{K}}_{\mu\nu}^{(\vartheta)}  =8\pi( T_{\mu\nu}^{(\vartheta)}+T_{\mu\nu}),\nn\\
\nabla_\mu\nabla^\mu\vartheta=\frac{\alpha}{4}e^\vartheta\mathcal{R}^2_{GB}
\label{eom.EdGB}
\eea
with $T_{\mu\nu}^{(\vartheta)} = \partial_\mu \vartheta \partial_\nu
\vartheta - \frac{1}{2}g_{\mu\nu} \partial_\sigma \vartheta \partial^\sigma \vartheta$ is the stress-energy tensor for the dilaton field.
The detailed form of ${\cal{K}}_{\mu\nu}^{(\vartheta)}$ can be found in \cite{Pani:2009wy}.

The waveforms for the \ac{EdGB} theory have been studied in many cases, such as the post-Newtonian waveforms from binary with comparable mass in a quasi-circular orbit \cite{Yagi:2011xp}.
This waveform is widely used in GW analysis. Based on this waveform and considering higher harmonics, more accurate waveforms were used to constrain EdGB \cite{Wang:2023wgv}. In this section, we construct the \ac{EMRI} waveform in the \ac{EdGB} theory in the case when the central \ac{MBH} is spinless.
The metric of spinless black holes in EdGB theory is founded in \cite{Yunes:2011we} as
\be
ds^2=-f_0[1+h_0(r)] dt^2+f_0^{-1}[1+k_0(r)] dr^2+r^2  d\Omega^2.
\label{MBH.metric}
\ee
The metric is obtained perturbatively. At the leading order,
\be
f_0=1-\frac{2M}r,
\ee
where $M$ is the black hole mass in \ac{GR}. By requiring the solution to be asymptotically flat and regular at $r=2 M$, one can find
\bea h_0&=&-\frac{49M}{40r}\zeta\Big(1+\frac{2 M}{r}+\frac{548}{147} \frac{M^2}{r^2}+\frac{8}{21} \frac{M^3}{r^3}\nn\\
&&\qquad\qquad -\frac{416}{147} \frac{M^4}{r^4}-\frac{1600}{147} \frac{M^{5}}{r^{5}}\Big)\,,\nn\\
k_0&=&\frac{49M}{40r}\zeta\Big(1+\frac{58}{49} \frac{M}{r}+\frac{76}{49} \frac{M^2}{r^2}-\frac{232}{21} \frac{M^3}{r^3} \nn\\
&&\qquad\qquad -\frac{3488}{147} \frac{M^4}{r^4}-\frac{7360}{147} \frac{M^{5}}{r^{5}}\Big)\,,\label{Et_hk}\eea
where $\zeta=16\pi\alpha^2/M^4$.
As we mentioned above, $\sqrt{\alpha}$ is constrained to be less than $\mathcal{O}(1)$km.
On the other hand, in our calculation for \acp{EMRI}, the mass of the \acp{MBH} $M$ will be about $10^6 M_\odot$, which corresponding to about $10^6$km.
Thus the value of $\zeta$ is as small as only $10^{-24}$, and thus the corresponding correction could be neglected in the waveform calculation,
and this is consistent with the result in \cite{Maselli:2020zgv}.
However, here we want to develop a self consistent waveform model, so we will still consider the modified metric and the corresponding geodesic in the following calculation.

\subsection{Geodesics}

For the central \ac{BH} as the spinless solution mentioned above,
we can choose the orbit of the \ac{SCO} on the equatorial plane due to the spherical symmetry,
which means that $\theta=\pi/2$, and $ d\theta/ d\tau=0$,
thus the Carter constant will also be 0 for equatorial orbit.
Then the geodesic equations can be written as:
\bea
\mu\frac{ dr}{ d\tau} &=& \pm\sqrt{\left(\frac{E^2}{f_0(1+h_0)}-\mu^2-\frac{L^2}{r^2}\right)\left(\frac{1-\frac{2M}{r}}{1+k_0}\right)},~~~ \label{OKN_v_r}\\
\mu\frac{ d\phi}{ d\tau} &=& \frac{L_z}{r^2}, \label{OKN_v_phi}\\
\mu\frac{ dt}{ d\tau} &=& \frac{E}{f_0(1+h_0)}.\label{OKN_v_t}
\eea

The radial coordinate $r$ will oscillation between the periastron $r_p$ and the apastron $r_a$, which makes the numerical integration become difficult to solve.
The periastron $r_p$ and the apastron $r_a$ corresponding to the point with $dr/d\tau=0$,
and thus the solution for the right-hand side of \eqref{OKN_v_r} equals to zero.
They can also be defined with the orbital eccentricity $e$ and semi-latus rectum $p$ as $r_p=p/(1+e)$ and $r_a=p/(1-e)$.
Then, we can define the angular variables $\psi$ as
\begin{equation}
 r=\frac{p}{1+e\cos\psi},
\label{OKN_rpsi}
\end{equation}
and thus the numerical integration of $\psi$ goes smoothly if the geodesic  equation of $r$ is transformed to that of $\psi$, and we have
\begin{eqnarray}
\frac{ d\psi}{ dt} &&= \frac{pf_0(1+h_0)}{Er^2e\sqrt{1-\cos^2\psi}} \times \nonumber \\
&&\sqrt{\left[\frac{E^2}{f_0(1+h_0)}-\mu^2-\frac{L^2}{r^2}\right]\left[\frac{1-2M/r}{1+k_0}\right]}~~~
\label{OKN_dpsidt}
\end{eqnarray}
the evolution of $\phi$ can also be written as
\be
\frac{d\phi}{dt}=\frac{L_zf_0(1+h_0)}{Er^2}
\ee

\subsection{Evolution of the constants}

We have introduced the geodesic for static \acp{BH} of the EdGB theory in the subsection above, but it's not the whole story of the trajectory.
With the emission of GW, the geodesic parameter will evolve due to the radiation reaction (RR).
The inspiral of the \acp{SCO} are driven by the RR effect, so different RR effects make different trajectories. In EdGB theory, the radiation of the dilaton will modify the RR effect, and the evolution of the geodesic will also be modified.
According to the study in \cite{Maselli:2020zgv}, this will dominated the difference between the waveform in EdGB and GR.

RR effect in the framework of EdGB inherits NK methods from GR, based on the evolution of the constants of motion $(E, L_z)$ that formulated in terms of 2PN order fluxes of energy and angular momentum in Teukolsky formalism \cite{Hughes:1999bq, Hughes:2001jr}, described by Gair and Glampedakis \cite{Gair:2005ih} in GHK formalization \cite{Glampedakis:2002cb}.
$\dot{E}$ and $\dot{L}$ can be decomposed into the GR part and the correction part from EdGB. For the GR term, we employ the result in \cite{Peters:1963ux,Peters:1964zz,Glampedakis:2002cb}:
\bea
\dot{E}_{GR}&=&\frac{-32\mu^2(1-e^2)^{3/2}}{5M^2}\left(\frac{M}{p}\right)^5\times\nn\\
&~&\left[\left(1+\frac{73}{24}e^2+\frac{37}{96}e^4\right)- \frac{a}{M}\left(\frac{M}{p}\right)^{3/2}\times\right.\nn\\
&~&\left.\left(\frac{73}{12}+
\frac{823}{24}e^2+\frac{949}{32}e^4+\frac{491}{192}e^6\right)\right]\\
\dot{L}_{z GR}&=&\frac{-32\mu^2 (1-e^2)^{3/2}}{5M}
\left (\frac{M}{p} \right )^{7/2}\left[\left(1+\frac{7}{8}e^2\right)\right.\nn\\
&~&-\left.\frac{a}{M}\left(\frac{M}{p}\right)^{3/2}\left(\frac{61}{12}+
\frac{119}{8}e^2+\frac{183}{32}e^4\right)\right]
\eea
One can see more detail in \cite{Gair:2005ih, Glampedakis:2002cb}.
In the EdGB case, the total energy flux and angular momentum flux  have the following structure:
\bea
\dot{E} &=& \dot{E}_{ GR} + \delta \dot{E} ,\\
\dot{L_z} &=& \dot{L_z}_{ GR} + \delta \dot{L_z} .
\eea
The correction $\delta\dot{E}$ and $\delta\dot{L_z}$ is dominated by the radiation of the dilaton field.
In the extreme mass-ratio limit, the fluxes of the dilaton field which dominated by the dipole radiation can be written as\cite{Loutrel:2014vja}:
\bea
\delta\dot{E}^{(\vartheta)} &=& -\frac{16\pi}{3} \alpha^2 \frac{\mu^2}{M^2p^4}(1-e^2)^{3/2}(1+\frac{e^2}{2}),~~~~ \\
\delta\dot{L_z}^{(\vartheta)} &=& -\frac{16\pi}{3} \alpha^2 \frac{\mu^2}{M^{5/2}p^{5/2}}(1-e^2)^{3/2}.
\eea

\subsection{Waveform Generation}
In this subsection, we will introduce the waveform generation of \ac{EMRI} in EdGB theory.
For the NK model, the waveform comes from the quadrupole radiation, which comes from the linearized  Einstein field equation.
The perturbative metric in the weak field can be decomposed into the Minkowski metric and a small perturbation as $g_{\mu\nu}=\eta_{\mu\nu}+h_{\mu\nu}$. With the Lorentz gauge condition, one can obtain the linearized Einstein equation
\begin{equation}
\square  \bar{h}^{\mu\nu} = -16\pi T^{\mu\nu}.
\label{WCD_liee}
\end{equation}
For the EdGB theory, We can find that the modified term only appear in the higher order terms, and thus the linearized equation is the same as that for GR in our calculation.
Under the approximations that the source is isolated, slowly moving, and far away from the observer,
one can impose the transverse-traceless (TT) gauge, and the solution of the linearized equation can be obtained as:
\begin{equation}
h^{jk}(t,\vec x) = \frac2r\left[\ddot{I}^{jk}(t-r)\right],
\label{OKN_quadwave}
\end{equation}
this is the waveform in the quadrupole formula. In which
\begin{equation}
I^{jk}(t^\prime) = \int x^{\prime j}x^{\prime k} T^{00}(t^\prime,x^\prime){ d}^3 x^\prime
\end{equation}
is the quadrupole moment tensor. In order to construct an ``equivalent flat space'' to provide a link between the trajectory and waveform generation, one can take a projection between Boyer-Lindquist coordinates onto a fictitious spherical polar coordinate grid in the Cartesian coordinate system. So that we can implement it in the trajectory constructed by geodesic and its correction.

\section{Response and Statistic}\label{sec:method}

\subsection{The Response of detector}

TianQin has a regular triangular configuration and is composed of three satellites centered around the earth at a distance of $L=\sqrt{3}\times10^8{\rm m}$.
For the detection of GWs, we need to consider the response of the detector. TianQin's three arms can be constructed into two independent interferometers.
Here we will not consider the time-delay interferometry,
and thus the signal $h_{I},  h_{II}$ for each channel can be written as:
\begin{equation}
h_{i}(t) = \frac{\sqrt{3}}{2}[F^+_i(t)h_+(t) + F_i^\times(t)h_\times(t)],
\end{equation}
$F^{+,\times}_i$ are the antenna pattern functions,
which are defined by $F_{\alpha_d}\equiv D^{ab}e^{+\times}_{ab}$\cite{Maggiore:1999vm, Apostolatos:1994mx}. Here $e^{+\times}_{ab}$ is the polarization tensor, and $D^{ab}\equiv \frac12 (u^a u^b - v^a v^b)$ is known as the detector tensor, which components $(u^i,v^j)$ determined by the orbit of TianQin satellites \cite{Hu:2018yqb, Fan:2020zhy, Feng:2019wgq}:
\begin{equation}
\begin{split}
x(t)=&R_e\cos\alpha_e+\frac{1}{2}R_e\cdot e_e\cdot \cos(2\alpha_e-3) \\
    &+\frac{1}{\sqrt{3}}L\cdot (\cos\theta_T\cos\phi_T\cos\gamma_e-\sin{\phi_T}\sin{\gamma_e})\,,\\	
y(t)=&R_e\sin\alpha_e+\frac{1}{2}R_e\cdot e_e\cdot \sin(2\alpha_e) \\
    &+\frac{1}{\sqrt{3}}L\cdot (\cos\theta_T\sin\phi_T\cos\gamma_e+\cos{\phi_T}\sin{\gamma_e})\,,\\
z(t)=&-\frac{1}{\sqrt{3}}L\cdot \sin\theta_T\cos\gamma_e\,.
\end{split}
\end{equation}
Where $R_e$ equals to 1AU, $e_e=0.0167$, and $\alpha_e=2\pi f_e t+\Phi_e$ is the phase comprised of the frequency $(f_e=1/{\rm yr})$ and initial phase of the Earth $(\Phi_e)$. $(\theta_T=-4.7^\circ, \phi_T=120.5^\circ)$ corresponding to the position of RX J0806.3+1527 which is the direction of the TianQIn. Denoting the phases of the satellites referring to their geocentric orbits, $\gamma_e$ has the expression $\gamma_e=2\pi t/T+2\pi n/3$, with $n=1,2,3$ corresponding to three satellites.

\subsection{Statistical method}

The inner product of the signals can be defined assuming the noise to be stationary and Gaussian with the power spectral density (PSD) of $S_N(f)$.
Thus the inner product between any two signals $(h(t),g(t))$ can be written as follows \cite{Cutler:1994ys, Balasubramanian:1995bm}:
\begin{equation}
(h|g) = 2\int^{\infty}_0 \frac{\tilde{h}(f)\tilde{g}^*(f) + \tilde{h}^*(f)\tilde{g}(f)}{S_N(f)}df.
\end{equation}
The noise PSD for TianQin is
\be
S_N(f)=\frac{1}{L^2}\left[\frac{S_a}{(2\pi f)^2}\left(1+\frac{10^{-4}Hz}{f}\right)+S_x\right]
\ee
with $S_a=10^{-30}m^2s^{-4}Hz^{-1}$, $S_x=10^{-24}m^2Hz^{-1}$,
And the \ac{SNR} for a specific signal $h$ can be written as
\begin{equation}
\rho = \sqrt{(h|h)}.
\end{equation}

To characterize the difference between two waveform, we can use the mismatch.
The mismatch between two waveforms $h_1$ and $h_2$ is defined as:
\be
\mathcal{M}(h_1,h_2) = 1-\frac{(h_1|h_2)}{\sqrt{(h_1|h_1)}\sqrt{(h_2|h_2)}}.
\ee
The mismatch will become 0 if $h_1=h_2$, and it will approaches to 1 for two waveforms that are completely different.
We can't distinguish two waveform models if the mismatch is small enough.
The threshold for the mismatch is \cite{Chatziioannou:2017tdw, Mangiagli:2018kpu, Baird:2012cu}:
\be \mathcal{M}_{th} = \frac{D}{2\rho^2}, \ee
where $D$ represents the dimension of the intrinsic parameters in the model, and it equals $6$ for our calculation, including semi-latus: $p$; eccentricity: $e$; MBH's mass: $M$; SCO's mass: $\mu$;  initial true anomaly: $\psi_0$; coupling constant $\alpha$. Other extrinsic parameters are: the position of the source in ecliptic coordinates: $(\theta_S,\phi_S)$, and the luminosity distance: $D$.

For the large \ac{SNR} case, the posterior of the parameters approx to a Gaussian distribution around the true value.
The the accuracy to estimate the parameters can be approximated as:
\begin{equation}
\Delta \lambda^i \simeq \sqrt{\Gamma_{ii}^{-1}}.
\end{equation}
$\Gamma_{ab}$ is the \ac{FIM} which is defined by
\begin{equation}
\Gamma_{ab} = \left(\frac{\partial  h}{\partial \lambda^a} \bigg | \frac{\partial  h}{\partial \lambda^b}\right).
\end{equation}
$\lambda_a$ are the parameters for the source.

\section{Constraints with TianQin}\label{sec:result}

\begin{figure}
\centering
\includegraphics[width=0.9\linewidth]{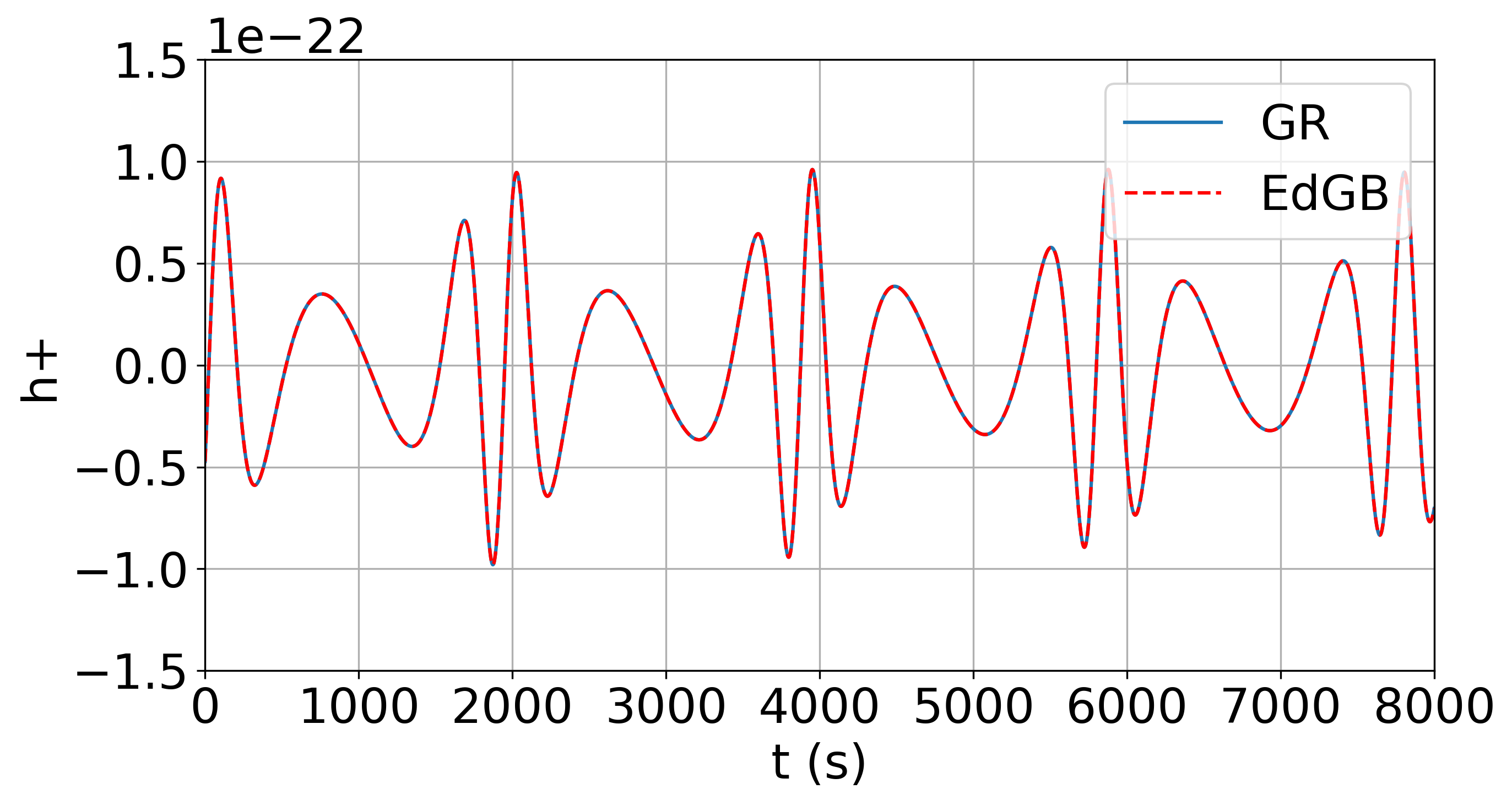}\\
\includegraphics[width=0.9\linewidth]{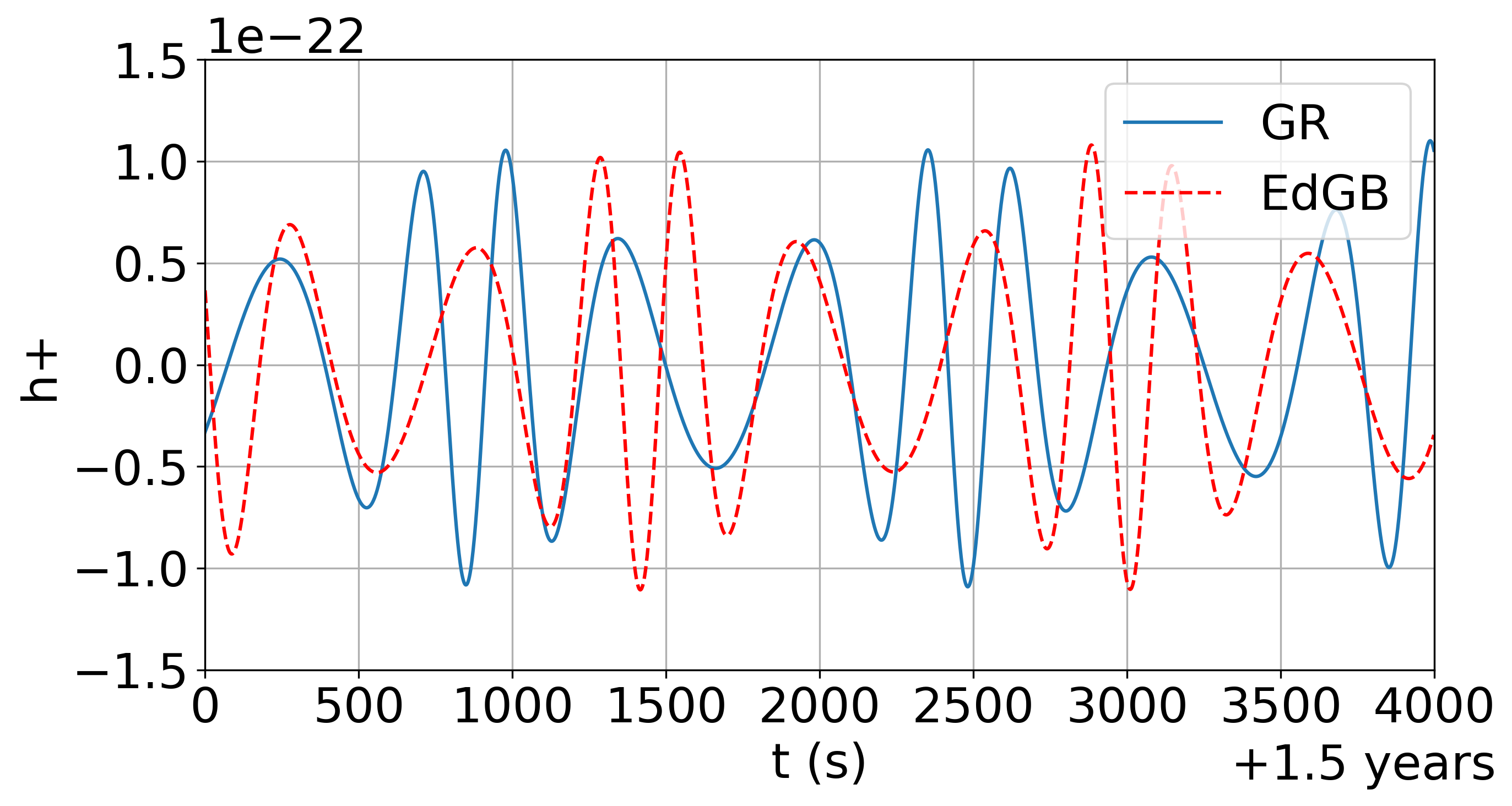}
\caption{Waveform of GR and EdGB with the same source parameters. The waveforms are displayed by $h_+$ polarization. The blue lines come from GR while the red lines are from EdGB. The top panel and the bottom panel illustrate the initial waveform and the waveform after 1.5 years, respectively.}
\label{figs_waveform}
\end{figure}

For the EMRI system with specific initial parameters, the waveform for GR and EdGB will be almost the same at the beginning.
Due to the difference on the evolution of the orbit, the waveforms will be different after a period of time.
As an example, we plot the waveform in Fig.\ref{figs_waveform} with the initial parameter as follows : $p=10.5M$; $e=0.4$; $\mu=10{ M_\odot}$; $M=10^6{ M_\odot}$;$\psi_0=0.1$; $D=1{\rm Gpc}$; $\theta_S=\phi_S=\pi/4$.
These are also the default parameters in the following calculation.
The coupling constant $\sqrt{\alpha}$ is 0 in GR and $\sqrt{10}{\rm km}$ in EdGB. One can notice that the waveform in GR and EdGB initially overlap completely, but there is a noticeable shift after 1.5 years. This shift has the potential to be captured by TianQin in the future.

\begin{figure}
\centering
\includegraphics[width=0.9\linewidth]{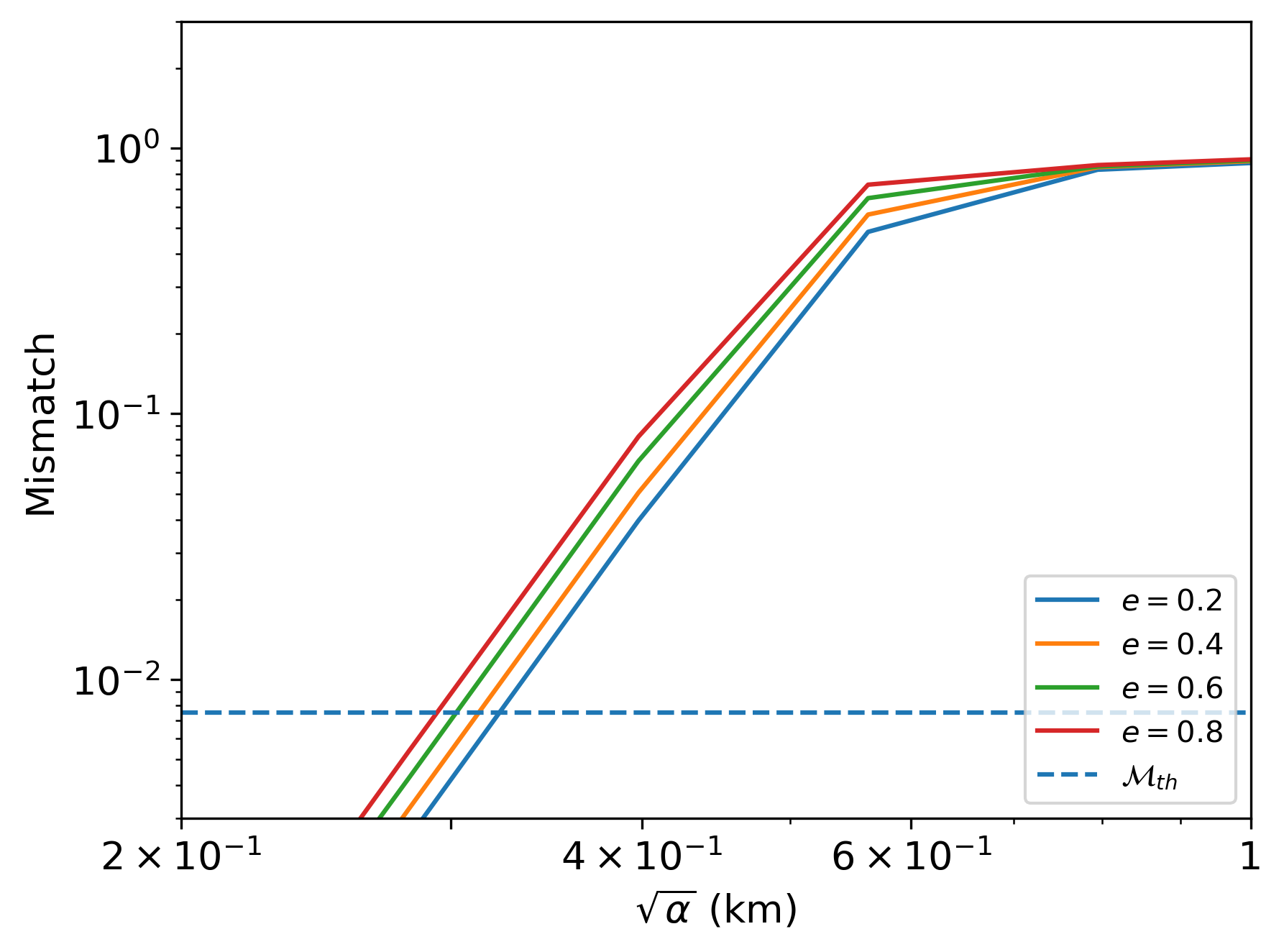}\\
\includegraphics[width=0.9\linewidth]{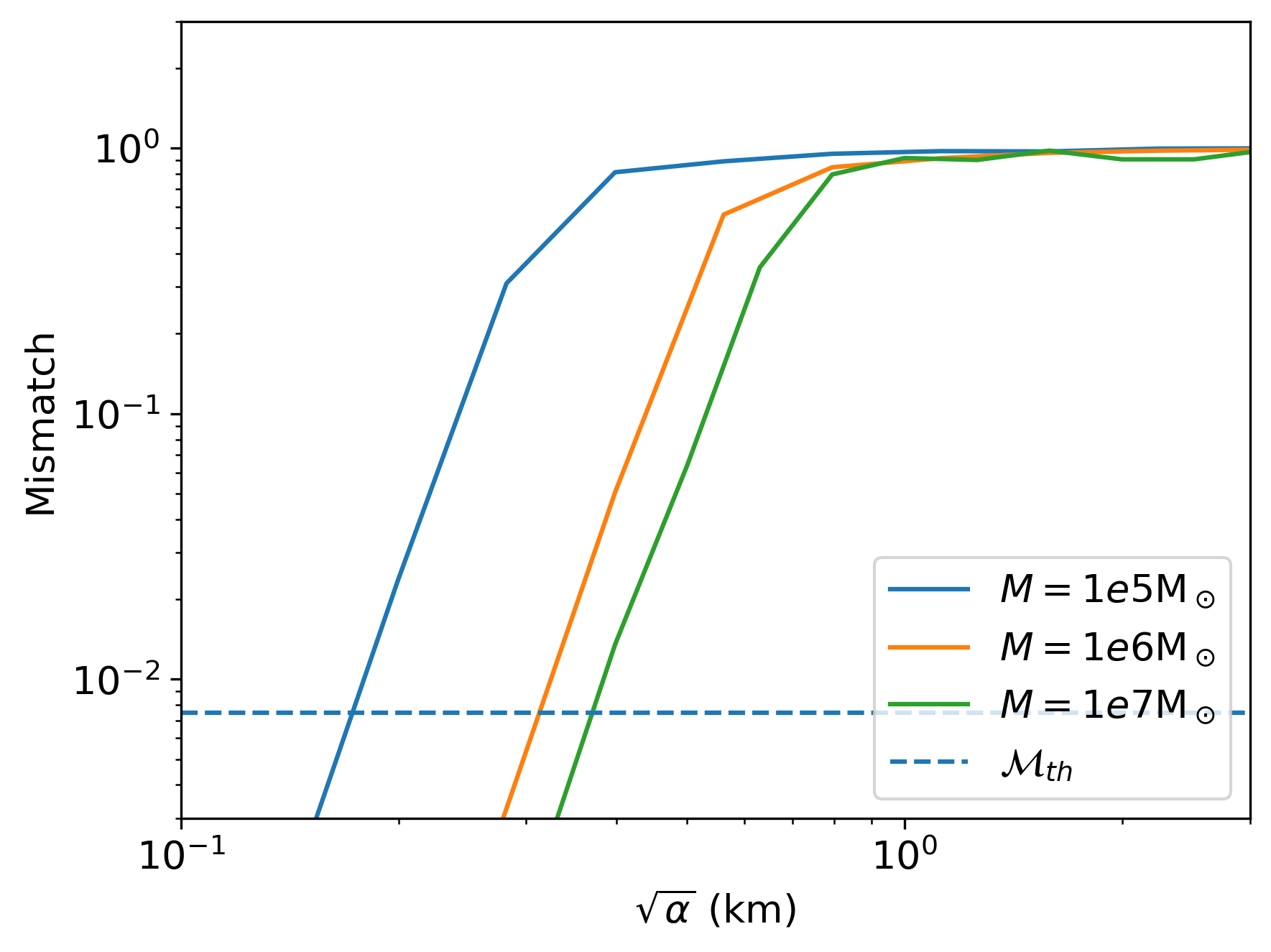}
\caption{Mismatch between GR and EdGB waveform. The top panel shows the result for the EMRIs with different eccentricitys. The Bottom panel shows the the result for the EMRIs with different mass of the central MBH.}
\label{figs_mismatch}
\end{figure}

The mismatch between GR and EdGB waveforms with one year of data is illustrated in Fig.\ref{figs_mismatch}.
With the detection threshold set to be \ac{SNR} = 20, the threshold of the mismatch is 0.0075.
All scenarios indicate that the mismatch increases with the increase of the EdGB coupling constant.
The mismatch tends toward 1 as long as the coupling parameter is sufficiently large; conversely, it approaches to 0.

On the top panel of Fig.\ref{figs_mismatch},
we fix all the other parameters except the eccentricity,
we can find that the mismatch approaches 1 more earlier with the increasing the coupling constant for larger eccentricity.
So the difference will become more sufficient for eccentric events,
but the influence is not significant since the lines are very close to each other.
On the bottom panel of Fig.\ref{figs_mismatch},
we fix all the other parameters except the mass of the MBH,
we can find that the mismatch approaches 1 more earlier with the increasing the coupling constant for larger mass,
and the influence is very significant.
This is caused by the fact that source with $M=10^5M_\odot$ is located at the most sensitive band of TianQin, and it should be worse for lighter sources.

Both results show that the threshold is passed when $\sqrt{\alpha}$ at the level of $\mathcal{O}(0.1)$km.
This indicates that the constraint may also approach to this level, which is improved by a order of magnitude on current results.

\begin{figure}
\centering
\includegraphics[width=0.9\linewidth]{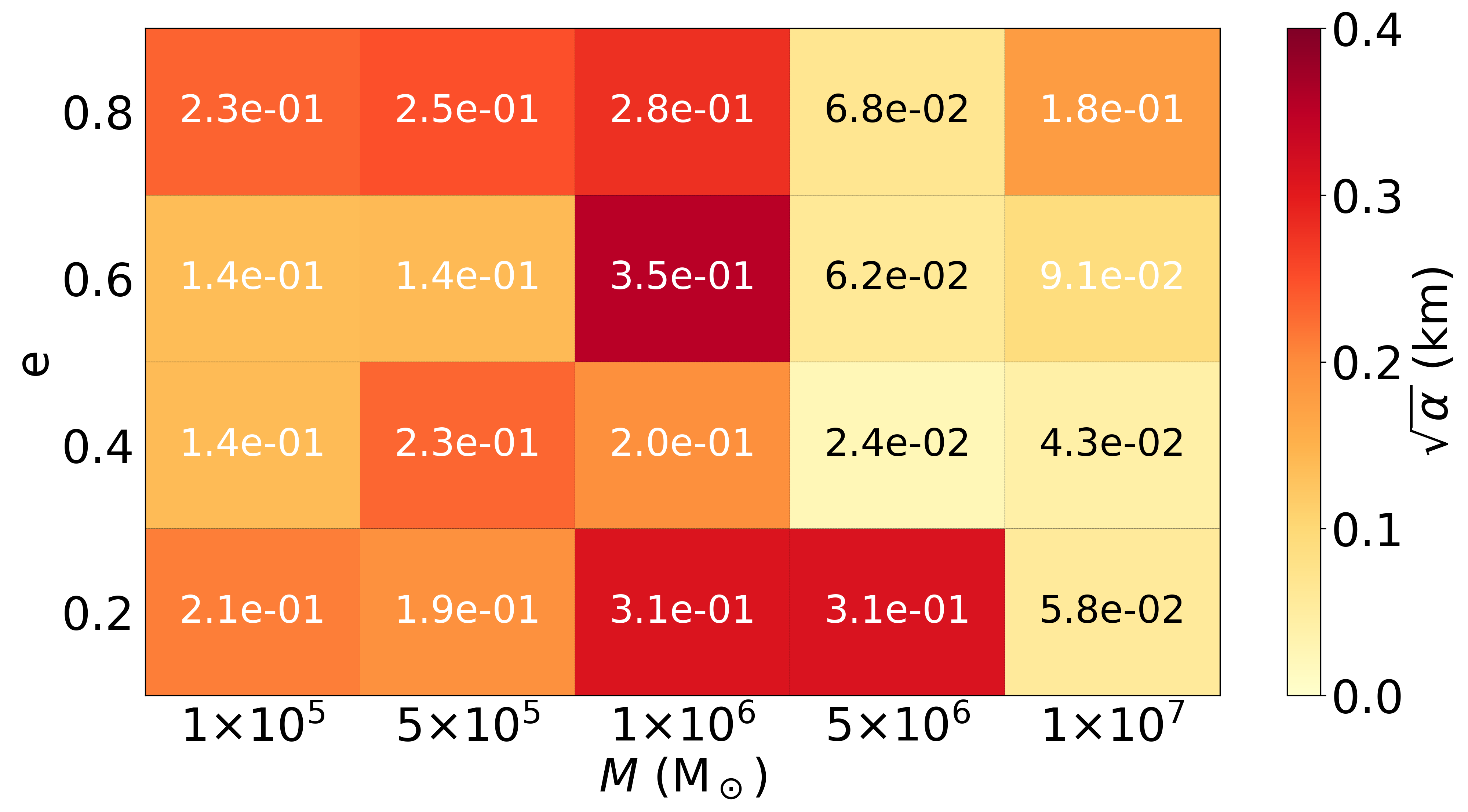}
\caption{Parameter estimation precisions for $\sqrt{\alpha}$ with TianQin. The results are calculated for the sources with different $M$ and $e$ with \ac{SNR} fixed to 50.
}
\label{figs_sqrt_alpha}
\end{figure}

For a more credible analysis, we also use the \ac{FIM} to calculate the capability of TianQin.
For a fair comparison, we fixed the SNR of the sources to be 50 by changing the luminosity distance $D$.
In Fig.\ref{figs_sqrt_alpha}, we demonstrates TianQin's precision on the \ac{PE} of the EdGB coupling parameter $\sqrt{\alpha}$ for different $M$ and $e$.
The result will varies for a few times for different sources, but all the precisions are at the order of $\mathcal{O}(0.1)$km.
This is consistent with the result of mismatch analysis, and the analysis within the PPE framework.

\section{Summary and Outlook}\label{sec:summary}

In this work, we construct the waveform of \acp{EMRI} in the \ac{EdGB} theory with the \ac{NK} method.
We use the static \ac{BH} solution in \ac{EdGB} and consider the equatorial eccentric orbit.
The geodesic equations is changed according to the modification of the metric.
Since the correction on the metric is relevant to $\zeta$,
which is as small as only $10^{-24}$ in our case for the \acp{MBH},
then its effect on the geodesic and the waveform is almost negligible.
In this work, we only include this term for a self consistent analysis.
On the other hand, due to the existence of the dilaton field, the additional dipole radiation will also cause the modification of the energy flux and the angular momentum flux.
This effect will dominant the modification of the waveform as the claim in previous studies.
Generally speaking, the linearized field equation also needs to be changed for an alternative theory of gravity.
However, for the case of \ac{EdGB} theory, the effect of this modification will not appear on the leading order corrections on the waveform.
Thus we will still use the ordinary quadrupole formula to generate the waveforms.

According to the generated waveform, we find that beginning from a common initial condition,
the waveforms for \ac{GR} and \ac{EdGB} which match with each other very well
will have significant differences after evolution for a period of time.
Then we carry out the mismatch analysis with the \ac{SNR} chosen to be 20, which is assumed to be the threshold for the detection of \acp{EMRI}.
The results show that the eccentricity will not have a significant impact on the result,
and the events with higher $e$ can distinguish two theories with smaller $\alpha$.
On the other hand, the mass of the \ac{MBH} will have more significant influence.
For all the cases, the mismatch can pass the threshold when the value of $\sqrt{\alpha}$ is about $0.1$ km.

We find that these results are consistent with the results of the \ac{FIM}.
The values of parameter measurement precision range from a few percents to one-tenth when fixing \ac{SNR}=50.

Since the central black hole we considered is non-spinning, and its orbits are confined to the equatorial, the waveforms we obtained is still limited.
In the future, if we consider a rotating black hole in EdGB, the orbits will extend into the three-dimensional space, and the result will be more realistic. Besides, the \ac{FIM} we used is also limited by the high SNR and the Gaussian posterior which may be violated for general cases.
The Bayesian inference may also be used in the future analysis.

\begin{acknowledgments}
The authors thank Yi-Ming Hu, Changfu Shi, Tieguang Zi, Shun-Jia Huang, Xue-Ting Zhang, and En-Kun Li for helpful discussions.
Jing Tan would also like to thank all the help from Chong-Bin Chen and Hong-Xiang Xu.
This work is supported by the Guangdong Basic and Applied Basic Research Foundation(Grant No. 2023A1515030116), the Guangdong Major Project of Basic and Applied Basic Research (Grant No. 2019B030302001), the National Key Research and Development Program of China (Grant No. 2021YFC2203002)and the National Science Foundation of China (Grant No. 12261131504).
\end{acknowledgments}
\bibliography{references}

\begin{thebibliography}{82}%
\makeatletter
\providecommand \@ifxundefined [1]{%
 \@ifx{#1\undefined}
}%
\providecommand \@ifnum [1]{%
 \ifnum #1\expandafter \@firstoftwo
 \else \expandafter \@secondoftwo
 \fi
}%
\providecommand \@ifx [1]{%
 \ifx #1\expandafter \@firstoftwo
 \else \expandafter \@secondoftwo
 \fi
}%
\providecommand \natexlab [1]{#1}%
\providecommand \enquote  [1]{``#1''}%
\providecommand \bibnamefont  [1]{#1}%
\providecommand \bibfnamefont [1]{#1}%
\providecommand \citenamefont [1]{#1}%
\providecommand \href@noop [0]{\@secondoftwo}%
\providecommand \href [0]{\begingroup \@sanitize@url \@href}%
\providecommand \@href[1]{\@@startlink{#1}\@@href}%
\providecommand \@@href[1]{\endgroup#1\@@endlink}%
\providecommand \@sanitize@url [0]{\catcode `\\12\catcode `\$12\catcode
  `\&12\catcode `\#12\catcode `\^12\catcode `\_12\catcode `\%12\relax}%
\providecommand \@@startlink[1]{}%
\providecommand \@@endlink[0]{}%
\providecommand \url  [0]{\begingroup\@sanitize@url \@url }%
\providecommand \@url [1]{\endgroup\@href {#1}{\urlprefix }}%
\providecommand \urlprefix  [0]{URL }%
\providecommand \Eprint [0]{\href }%
\providecommand \doibase [0]{http://dx.doi.org/}%
\providecommand \selectlanguage [0]{\@gobble}%
\providecommand \bibinfo  [0]{\@secondoftwo}%
\providecommand \bibfield  [0]{\@secondoftwo}%
\providecommand \translation [1]{[#1]}%
\providecommand \BibitemOpen [0]{}%
\providecommand \bibitemStop [0]{}%
\providecommand \bibitemNoStop [0]{.\EOS\space}%
\providecommand \EOS [0]{\spacefactor3000\relax}%
\providecommand \BibitemShut  [1]{\csname bibitem#1\endcsname}%
\let\auto@bib@innerbib\@empty
\bibitem [{\citenamefont {Luo}\ \emph {et~al.}(2016)\citenamefont {Luo} \emph
  {et~al.}}]{Luo:2015ght}%
  \BibitemOpen
  \bibfield  {author} {\bibinfo {author} {\bibfnamefont {J.}~\bibnamefont
  {Luo}} \emph {et~al.} (\bibinfo {collaboration} {TianQin}),\ }\href {\doibase
  10.1088/0264-9381/33/3/035010} {\bibfield  {journal} {\bibinfo  {journal}
  {Class. Quant. Grav.}\ }\textbf {\bibinfo {volume} {33}},\ \bibinfo {pages}
  {035010} (\bibinfo {year} {2016})},\ \Eprint
  {http://arxiv.org/abs/1512.02076} {arXiv:1512.02076 [astro-ph.IM]}
  \BibitemShut {NoStop}%
\bibitem [{\citenamefont {Mei}\ \emph {et~al.}(2021)\citenamefont {Mei} \emph
  {et~al.}}]{TianQin:2020hid}%
  \BibitemOpen
  \bibfield  {author} {\bibinfo {author} {\bibfnamefont {J.}~\bibnamefont
  {Mei}} \emph {et~al.} (\bibinfo {collaboration} {TianQin}),\ }\href {\doibase
  10.1093/ptep/ptaa114} {\bibfield  {journal} {\bibinfo  {journal} {PTEP}\
  }\textbf {\bibinfo {volume} {2021}},\ \bibinfo {pages} {05A107} (\bibinfo
  {year} {2021})},\ \Eprint {http://arxiv.org/abs/2008.10332} {arXiv:2008.10332
  [gr-qc]} \BibitemShut {NoStop}%
\bibitem [{\citenamefont {Danzmann}(1997)}]{Danzmann:1997hm}%
  \BibitemOpen
  \bibfield  {author} {\bibinfo {author} {\bibfnamefont {K.}~\bibnamefont
  {Danzmann}},\ }\href {\doibase 10.1088/0264-9381/14/6/002} {\bibfield
  {journal} {\bibinfo  {journal} {Class. Quant. Grav.}\ }\textbf {\bibinfo
  {volume} {14}},\ \bibinfo {pages} {1399} (\bibinfo {year}
  {1997})}\BibitemShut {NoStop}%
\bibitem [{\citenamefont {Amaro-Seoane}\ \emph {et~al.}(2017)\citenamefont
  {Amaro-Seoane} \emph {et~al.}}]{LISA:2017pwj}%
  \BibitemOpen
  \bibfield  {author} {\bibinfo {author} {\bibfnamefont {P.}~\bibnamefont
  {Amaro-Seoane}} \emph {et~al.} (\bibinfo {collaboration} {LISA}),\
  }\href@noop {} {\  (\bibinfo {year} {2017})},\ \Eprint
  {http://arxiv.org/abs/1702.00786} {arXiv:1702.00786 [astro-ph.IM]}
  \BibitemShut {NoStop}%
\bibitem [{\citenamefont {Hu}\ and\ \citenamefont {Wu}(2017)}]{Hu:2017mde}%
  \BibitemOpen
  \bibfield  {author} {\bibinfo {author} {\bibfnamefont {W.-R.}\ \bibnamefont
  {Hu}}\ and\ \bibinfo {author} {\bibfnamefont {Y.-L.}\ \bibnamefont {Wu}},\
  }\href {\doibase 10.1093/nsr/nwx116} {\bibfield  {journal} {\bibinfo
  {journal} {Natl. Sci. Rev.}\ }\textbf {\bibinfo {volume} {4}},\ \bibinfo
  {pages} {685} (\bibinfo {year} {2017})}\BibitemShut {NoStop}%
\bibitem [{\citenamefont {Korol}\ \emph {et~al.}(2017)\citenamefont {Korol},
  \citenamefont {Rossi}, \citenamefont {Groot}, \citenamefont {Nelemans},
  \citenamefont {Toonen},\ and\ \citenamefont {Brown}}]{Korol:2017qcx}%
  \BibitemOpen
  \bibfield  {author} {\bibinfo {author} {\bibfnamefont {V.}~\bibnamefont
  {Korol}}, \bibinfo {author} {\bibfnamefont {E.~M.}\ \bibnamefont {Rossi}},
  \bibinfo {author} {\bibfnamefont {P.~J.}\ \bibnamefont {Groot}}, \bibinfo
  {author} {\bibfnamefont {G.}~\bibnamefont {Nelemans}}, \bibinfo {author}
  {\bibfnamefont {S.}~\bibnamefont {Toonen}}, \ and\ \bibinfo {author}
  {\bibfnamefont {A.~G.~A.}\ \bibnamefont {Brown}},\ }\href {\doibase
  10.1093/mnras/stx1285} {\bibfield  {journal} {\bibinfo  {journal} {Mon. Not.
  Roy. Astron. Soc.}\ }\textbf {\bibinfo {volume} {470}},\ \bibinfo {pages}
  {1894} (\bibinfo {year} {2017})},\ \Eprint {http://arxiv.org/abs/1703.02555}
  {arXiv:1703.02555 [astro-ph.HE]} \BibitemShut {NoStop}%
\bibitem [{\citenamefont {Huang}\ \emph {et~al.}(2020)\citenamefont {Huang},
  \citenamefont {Hu}, \citenamefont {Korol}, \citenamefont {Li}, \citenamefont
  {Liang}, \citenamefont {Lu}, \citenamefont {Wang}, \citenamefont {Yu},\ and\
  \citenamefont {Mei}}]{Huang:2020rjf}%
  \BibitemOpen
  \bibfield  {author} {\bibinfo {author} {\bibfnamefont {S.-J.}\ \bibnamefont
  {Huang}}, \bibinfo {author} {\bibfnamefont {Y.-M.}\ \bibnamefont {Hu}},
  \bibinfo {author} {\bibfnamefont {V.}~\bibnamefont {Korol}}, \bibinfo
  {author} {\bibfnamefont {P.-C.}\ \bibnamefont {Li}}, \bibinfo {author}
  {\bibfnamefont {Z.-C.}\ \bibnamefont {Liang}}, \bibinfo {author}
  {\bibfnamefont {Y.}~\bibnamefont {Lu}}, \bibinfo {author} {\bibfnamefont
  {H.-T.}\ \bibnamefont {Wang}}, \bibinfo {author} {\bibfnamefont
  {S.}~\bibnamefont {Yu}}, \ and\ \bibinfo {author} {\bibfnamefont
  {J.}~\bibnamefont {Mei}},\ }\href {\doibase 10.1103/PhysRevD.102.063021}
  {\bibfield  {journal} {\bibinfo  {journal} {Phys. Rev. D}\ }\textbf {\bibinfo
  {volume} {102}},\ \bibinfo {pages} {063021} (\bibinfo {year} {2020})},\
  \Eprint {http://arxiv.org/abs/2005.07889} {arXiv:2005.07889 [astro-ph.HE]}
  \BibitemShut {NoStop}%
\bibitem [{\citenamefont {Klein}\ \emph {et~al.}(2016)\citenamefont {Klein}
  \emph {et~al.}}]{Klein:2015hvg}%
  \BibitemOpen
  \bibfield  {author} {\bibinfo {author} {\bibfnamefont {A.}~\bibnamefont
  {Klein}} \emph {et~al.},\ }\href {\doibase 10.1103/PhysRevD.93.024003}
  {\bibfield  {journal} {\bibinfo  {journal} {Phys. Rev. D}\ }\textbf {\bibinfo
  {volume} {93}},\ \bibinfo {pages} {024003} (\bibinfo {year} {2016})},\
  \Eprint {http://arxiv.org/abs/1511.05581} {arXiv:1511.05581 [gr-qc]}
  \BibitemShut {NoStop}%
\bibitem [{\citenamefont {Wang}\ \emph {et~al.}(2019)\citenamefont {Wang} \emph
  {et~al.}}]{Wang:2019ryf}%
  \BibitemOpen
  \bibfield  {author} {\bibinfo {author} {\bibfnamefont {H.-T.}\ \bibnamefont
  {Wang}} \emph {et~al.},\ }\href {\doibase 10.1103/PhysRevD.100.043003}
  {\bibfield  {journal} {\bibinfo  {journal} {Phys. Rev. D}\ }\textbf {\bibinfo
  {volume} {100}},\ \bibinfo {pages} {043003} (\bibinfo {year} {2019})},\
  \Eprint {http://arxiv.org/abs/1902.04423} {arXiv:1902.04423 [astro-ph.HE]}
  \BibitemShut {NoStop}%
\bibitem [{\citenamefont {Feng}\ \emph {et~al.}(2019)\citenamefont {Feng},
  \citenamefont {Wang}, \citenamefont {Hu}, \citenamefont {Hu},\ and\
  \citenamefont {Wang}}]{Feng:2019wgq}%
  \BibitemOpen
  \bibfield  {author} {\bibinfo {author} {\bibfnamefont {W.-F.}\ \bibnamefont
  {Feng}}, \bibinfo {author} {\bibfnamefont {H.-T.}\ \bibnamefont {Wang}},
  \bibinfo {author} {\bibfnamefont {X.-C.}\ \bibnamefont {Hu}}, \bibinfo
  {author} {\bibfnamefont {Y.-M.}\ \bibnamefont {Hu}}, \ and\ \bibinfo {author}
  {\bibfnamefont {Y.}~\bibnamefont {Wang}},\ }\href {\doibase
  10.1103/PhysRevD.99.123002} {\bibfield  {journal} {\bibinfo  {journal} {Phys.
  Rev. D}\ }\textbf {\bibinfo {volume} {99}},\ \bibinfo {pages} {123002}
  (\bibinfo {year} {2019})},\ \Eprint {http://arxiv.org/abs/1901.02159}
  {arXiv:1901.02159 [astro-ph.IM]} \BibitemShut {NoStop}%
\bibitem [{\citenamefont {Sesana}(2016)}]{Sesana:2016ljz}%
  \BibitemOpen
  \bibfield  {author} {\bibinfo {author} {\bibfnamefont {A.}~\bibnamefont
  {Sesana}},\ }\href {\doibase 10.1103/PhysRevLett.116.231102} {\bibfield
  {journal} {\bibinfo  {journal} {Phys. Rev. Lett.}\ }\textbf {\bibinfo
  {volume} {116}},\ \bibinfo {pages} {231102} (\bibinfo {year} {2016})},\
  \Eprint {http://arxiv.org/abs/1602.06951} {arXiv:1602.06951 [gr-qc]}
  \BibitemShut {NoStop}%
\bibitem [{\citenamefont {Kyutoku}\ and\ \citenamefont
  {Seto}(2016)}]{Kyutoku:2016ppx}%
  \BibitemOpen
  \bibfield  {author} {\bibinfo {author} {\bibfnamefont {K.}~\bibnamefont
  {Kyutoku}}\ and\ \bibinfo {author} {\bibfnamefont {N.}~\bibnamefont {Seto}},\
  }\href {\doibase 10.1093/mnras/stw1767} {\bibfield  {journal} {\bibinfo
  {journal} {Mon. Not. Roy. Astron. Soc.}\ }\textbf {\bibinfo {volume} {462}},\
  \bibinfo {pages} {2177} (\bibinfo {year} {2016})},\ \Eprint
  {http://arxiv.org/abs/1606.02298} {arXiv:1606.02298 [astro-ph.HE]}
  \BibitemShut {NoStop}%
\bibitem [{\citenamefont {Liu}\ \emph {et~al.}(2020)\citenamefont {Liu},
  \citenamefont {Hu}, \citenamefont {Zhang},\ and\ \citenamefont
  {Mei}}]{Liu:2020eko}%
  \BibitemOpen
  \bibfield  {author} {\bibinfo {author} {\bibfnamefont {S.}~\bibnamefont
  {Liu}}, \bibinfo {author} {\bibfnamefont {Y.-M.}\ \bibnamefont {Hu}},
  \bibinfo {author} {\bibfnamefont {J.-d.}\ \bibnamefont {Zhang}}, \ and\
  \bibinfo {author} {\bibfnamefont {J.}~\bibnamefont {Mei}},\ }\href {\doibase
  10.1103/PhysRevD.101.103027} {\bibfield  {journal} {\bibinfo  {journal}
  {Phys. Rev. D}\ }\textbf {\bibinfo {volume} {101}},\ \bibinfo {pages}
  {103027} (\bibinfo {year} {2020})},\ \Eprint
  {http://arxiv.org/abs/2004.14242} {arXiv:2004.14242 [astro-ph.HE]}
  \BibitemShut {NoStop}%
\bibitem [{\citenamefont {Babak}\ \emph {et~al.}(2017)\citenamefont {Babak},
  \citenamefont {Gair}, \citenamefont {Sesana}, \citenamefont {Barausse},
  \citenamefont {Sopuerta}, \citenamefont {Berry}, \citenamefont {Berti},
  \citenamefont {Amaro-Seoane}, \citenamefont {Petiteau},\ and\ \citenamefont
  {Klein}}]{Babak:2017tow}%
  \BibitemOpen
  \bibfield  {author} {\bibinfo {author} {\bibfnamefont {S.}~\bibnamefont
  {Babak}}, \bibinfo {author} {\bibfnamefont {J.}~\bibnamefont {Gair}},
  \bibinfo {author} {\bibfnamefont {A.}~\bibnamefont {Sesana}}, \bibinfo
  {author} {\bibfnamefont {E.}~\bibnamefont {Barausse}}, \bibinfo {author}
  {\bibfnamefont {C.~F.}\ \bibnamefont {Sopuerta}}, \bibinfo {author}
  {\bibfnamefont {C.~P.~L.}\ \bibnamefont {Berry}}, \bibinfo {author}
  {\bibfnamefont {E.}~\bibnamefont {Berti}}, \bibinfo {author} {\bibfnamefont
  {P.}~\bibnamefont {Amaro-Seoane}}, \bibinfo {author} {\bibfnamefont
  {A.}~\bibnamefont {Petiteau}}, \ and\ \bibinfo {author} {\bibfnamefont
  {A.}~\bibnamefont {Klein}},\ }\href {\doibase 10.1103/PhysRevD.95.103012}
  {\bibfield  {journal} {\bibinfo  {journal} {Phys. Rev. D}\ }\textbf {\bibinfo
  {volume} {95}},\ \bibinfo {pages} {103012} (\bibinfo {year} {2017})},\
  \Eprint {http://arxiv.org/abs/1703.09722} {arXiv:1703.09722 [gr-qc]}
  \BibitemShut {NoStop}%
\bibitem [{\citenamefont {Fan}\ \emph {et~al.}(2020)\citenamefont {Fan},
  \citenamefont {Hu}, \citenamefont {Barausse}, \citenamefont {Sesana},
  \citenamefont {Zhang}, \citenamefont {Zhang}, \citenamefont {Zi},\ and\
  \citenamefont {Mei}}]{Fan:2020zhy}%
  \BibitemOpen
  \bibfield  {author} {\bibinfo {author} {\bibfnamefont {H.-M.}\ \bibnamefont
  {Fan}}, \bibinfo {author} {\bibfnamefont {Y.-M.}\ \bibnamefont {Hu}},
  \bibinfo {author} {\bibfnamefont {E.}~\bibnamefont {Barausse}}, \bibinfo
  {author} {\bibfnamefont {A.}~\bibnamefont {Sesana}}, \bibinfo {author}
  {\bibfnamefont {J.-d.}\ \bibnamefont {Zhang}}, \bibinfo {author}
  {\bibfnamefont {X.}~\bibnamefont {Zhang}}, \bibinfo {author} {\bibfnamefont
  {T.-G.}\ \bibnamefont {Zi}}, \ and\ \bibinfo {author} {\bibfnamefont
  {J.}~\bibnamefont {Mei}},\ }\href {\doibase 10.1103/PhysRevD.102.063016}
  {\bibfield  {journal} {\bibinfo  {journal} {Phys. Rev. D}\ }\textbf {\bibinfo
  {volume} {102}},\ \bibinfo {pages} {063016} (\bibinfo {year} {2020})},\
  \Eprint {http://arxiv.org/abs/2005.08212} {arXiv:2005.08212 [astro-ph.HE]}
  \BibitemShut {NoStop}%
\bibitem [{\citenamefont {Caprini}\ \emph {et~al.}(2016)\citenamefont {Caprini}
  \emph {et~al.}}]{Caprini:2015zlo}%
  \BibitemOpen
  \bibfield  {author} {\bibinfo {author} {\bibfnamefont {C.}~\bibnamefont
  {Caprini}} \emph {et~al.},\ }\href {\doibase 10.1088/1475-7516/2016/04/001}
  {\bibfield  {journal} {\bibinfo  {journal} {JCAP}\ }\textbf {\bibinfo
  {volume} {04}},\ \bibinfo {pages} {001} (\bibinfo {year} {2016})},\ \Eprint
  {http://arxiv.org/abs/1512.06239} {arXiv:1512.06239 [astro-ph.CO]}
  \BibitemShut {NoStop}%
\bibitem [{\citenamefont {Bartolo}\ \emph {et~al.}(2016)\citenamefont {Bartolo}
  \emph {et~al.}}]{Bartolo:2016ami}%
  \BibitemOpen
  \bibfield  {author} {\bibinfo {author} {\bibfnamefont {N.}~\bibnamefont
  {Bartolo}} \emph {et~al.},\ }\href {\doibase 10.1088/1475-7516/2016/12/026}
  {\bibfield  {journal} {\bibinfo  {journal} {JCAP}\ }\textbf {\bibinfo
  {volume} {12}},\ \bibinfo {pages} {026} (\bibinfo {year} {2016})},\ \Eprint
  {http://arxiv.org/abs/1610.06481} {arXiv:1610.06481 [astro-ph.CO]}
  \BibitemShut {NoStop}%
\bibitem [{\citenamefont {Liang}\ \emph {et~al.}(2022)\citenamefont {Liang},
  \citenamefont {Hu}, \citenamefont {Jiang}, \citenamefont {Cheng},
  \citenamefont {Zhang},\ and\ \citenamefont {Mei}}]{Liang:2021bde}%
  \BibitemOpen
  \bibfield  {author} {\bibinfo {author} {\bibfnamefont {Z.-C.}\ \bibnamefont
  {Liang}}, \bibinfo {author} {\bibfnamefont {Y.-M.}\ \bibnamefont {Hu}},
  \bibinfo {author} {\bibfnamefont {Y.}~\bibnamefont {Jiang}}, \bibinfo
  {author} {\bibfnamefont {J.}~\bibnamefont {Cheng}}, \bibinfo {author}
  {\bibfnamefont {J.-d.}\ \bibnamefont {Zhang}}, \ and\ \bibinfo {author}
  {\bibfnamefont {J.}~\bibnamefont {Mei}},\ }\href {\doibase
  10.1103/PhysRevD.105.022001} {\bibfield  {journal} {\bibinfo  {journal}
  {Phys. Rev. D}\ }\textbf {\bibinfo {volume} {105}},\ \bibinfo {pages}
  {022001} (\bibinfo {year} {2022})},\ \Eprint
  {http://arxiv.org/abs/2107.08643} {arXiv:2107.08643 [astro-ph.CO]}
  \BibitemShut {NoStop}%
\bibitem [{\citenamefont {Cheng}\ \emph {et~al.}(2022)\citenamefont {Cheng},
  \citenamefont {Li}, \citenamefont {Hu}, \citenamefont {Liang}, \citenamefont
  {Zhang},\ and\ \citenamefont {Mei}}]{Cheng:2022vct}%
  \BibitemOpen
  \bibfield  {author} {\bibinfo {author} {\bibfnamefont {J.}~\bibnamefont
  {Cheng}}, \bibinfo {author} {\bibfnamefont {E.-K.}\ \bibnamefont {Li}},
  \bibinfo {author} {\bibfnamefont {Y.-M.}\ \bibnamefont {Hu}}, \bibinfo
  {author} {\bibfnamefont {Z.-C.}\ \bibnamefont {Liang}}, \bibinfo {author}
  {\bibfnamefont {J.-d.}\ \bibnamefont {Zhang}}, \ and\ \bibinfo {author}
  {\bibfnamefont {J.}~\bibnamefont {Mei}},\ }\href {\doibase
  10.1103/PhysRevD.106.124027} {\bibfield  {journal} {\bibinfo  {journal}
  {Phys. Rev. D}\ }\textbf {\bibinfo {volume} {106}},\ \bibinfo {pages}
  {124027} (\bibinfo {year} {2022})},\ \Eprint
  {http://arxiv.org/abs/2208.11615} {arXiv:2208.11615 [gr-qc]} \BibitemShut
  {NoStop}%
\bibitem [{\citenamefont {Gair}\ \emph {et~al.}(2013)\citenamefont {Gair},
  \citenamefont {Vallisneri}, \citenamefont {Larson},\ and\ \citenamefont
  {Baker}}]{Gair:2012nm}%
  \BibitemOpen
  \bibfield  {author} {\bibinfo {author} {\bibfnamefont {J.~R.}\ \bibnamefont
  {Gair}}, \bibinfo {author} {\bibfnamefont {M.}~\bibnamefont {Vallisneri}},
  \bibinfo {author} {\bibfnamefont {S.~L.}\ \bibnamefont {Larson}}, \ and\
  \bibinfo {author} {\bibfnamefont {J.~G.}\ \bibnamefont {Baker}},\ }\href
  {\doibase 10.12942/lrr-2013-7} {\bibfield  {journal} {\bibinfo  {journal}
  {Living Rev. Rel.}\ }\textbf {\bibinfo {volume} {16}},\ \bibinfo {pages} {7}
  (\bibinfo {year} {2013})},\ \Eprint {http://arxiv.org/abs/1212.5575}
  {arXiv:1212.5575 [gr-qc]} \BibitemShut {NoStop}%
\bibitem [{\citenamefont {Barausse}\ \emph {et~al.}(2020)\citenamefont
  {Barausse} \emph {et~al.}}]{Barausse:2020rsu}%
  \BibitemOpen
  \bibfield  {author} {\bibinfo {author} {\bibfnamefont {E.}~\bibnamefont
  {Barausse}} \emph {et~al.},\ }\href {\doibase 10.1007/s10714-020-02691-1}
  {\bibfield  {journal} {\bibinfo  {journal} {Gen. Rel. Grav.}\ }\textbf
  {\bibinfo {volume} {52}},\ \bibinfo {pages} {81} (\bibinfo {year} {2020})},\
  \Eprint {http://arxiv.org/abs/2001.09793} {arXiv:2001.09793 [gr-qc]}
  \BibitemShut {NoStop}%
\bibitem [{\citenamefont {Arun}\ \emph {et~al.}(2022)\citenamefont {Arun} \emph
  {et~al.}}]{LISA:2022kgy}%
  \BibitemOpen
  \bibfield  {author} {\bibinfo {author} {\bibfnamefont {K.~G.}\ \bibnamefont
  {Arun}} \emph {et~al.} (\bibinfo {collaboration} {LISA}),\ }\href {\doibase
  10.1007/s41114-022-00036-9} {\bibfield  {journal} {\bibinfo  {journal}
  {Living Rev. Rel.}\ }\textbf {\bibinfo {volume} {25}},\ \bibinfo {pages} {4}
  (\bibinfo {year} {2022})},\ \Eprint {http://arxiv.org/abs/2205.01597}
  {arXiv:2205.01597 [gr-qc]} \BibitemShut {NoStop}%
\bibitem [{\citenamefont {Seoane}\ \emph {et~al.}(2023)\citenamefont {Seoane}
  \emph {et~al.}}]{LISA:2022yao}%
  \BibitemOpen
  \bibfield  {author} {\bibinfo {author} {\bibfnamefont {P.~A.}\ \bibnamefont
  {Seoane}} \emph {et~al.} (\bibinfo {collaboration} {LISA}),\ }\href {\doibase
  10.1007/s41114-022-00041-y} {\bibfield  {journal} {\bibinfo  {journal}
  {Living Rev. Rel.}\ }\textbf {\bibinfo {volume} {26}},\ \bibinfo {pages} {2}
  (\bibinfo {year} {2023})},\ \Eprint {http://arxiv.org/abs/2203.06016}
  {arXiv:2203.06016 [gr-qc]} \BibitemShut {NoStop}%
\bibitem [{\citenamefont {Baker}\ \emph {et~al.}(2019)\citenamefont {Baker}
  \emph {et~al.}}]{Baker:2019pnp}%
  \BibitemOpen
  \bibfield  {author} {\bibinfo {author} {\bibfnamefont {J.}~\bibnamefont
  {Baker}} \emph {et~al.},\ }\href@noop {} {\bibfield  {journal} {\bibinfo
  {journal} {Bull. Am. Astron. Soc.}\ }\textbf {\bibinfo {volume} {51}},\
  \bibinfo {pages} {243} (\bibinfo {year} {2019})},\ \Eprint
  {http://arxiv.org/abs/1907.11305} {arXiv:1907.11305 [astro-ph.IM]}
  \BibitemShut {NoStop}%
\bibitem [{\citenamefont {Tamanini}\ \emph {et~al.}(2016)\citenamefont
  {Tamanini}, \citenamefont {Caprini}, \citenamefont {Barausse}, \citenamefont
  {Sesana}, \citenamefont {Klein},\ and\ \citenamefont
  {Petiteau}}]{Tamanini:2016zlh}%
  \BibitemOpen
  \bibfield  {author} {\bibinfo {author} {\bibfnamefont {N.}~\bibnamefont
  {Tamanini}}, \bibinfo {author} {\bibfnamefont {C.}~\bibnamefont {Caprini}},
  \bibinfo {author} {\bibfnamefont {E.}~\bibnamefont {Barausse}}, \bibinfo
  {author} {\bibfnamefont {A.}~\bibnamefont {Sesana}}, \bibinfo {author}
  {\bibfnamefont {A.}~\bibnamefont {Klein}}, \ and\ \bibinfo {author}
  {\bibfnamefont {A.}~\bibnamefont {Petiteau}},\ }\href {\doibase
  10.1088/1475-7516/2016/04/002} {\bibfield  {journal} {\bibinfo  {journal}
  {JCAP}\ }\textbf {\bibinfo {volume} {04}},\ \bibinfo {pages} {002} (\bibinfo
  {year} {2016})},\ \Eprint {http://arxiv.org/abs/1601.07112} {arXiv:1601.07112
  [astro-ph.CO]} \BibitemShut {NoStop}%
\bibitem [{\citenamefont {Zhu}\ \emph {et~al.}(2022)\citenamefont {Zhu},
  \citenamefont {Hu}, \citenamefont {Wang}, \citenamefont {Zhang},
  \citenamefont {Li}, \citenamefont {Hendry},\ and\ \citenamefont
  {Mei}}]{Zhu:2021aat}%
  \BibitemOpen
  \bibfield  {author} {\bibinfo {author} {\bibfnamefont {L.-G.}\ \bibnamefont
  {Zhu}}, \bibinfo {author} {\bibfnamefont {Y.-M.}\ \bibnamefont {Hu}},
  \bibinfo {author} {\bibfnamefont {H.-T.}\ \bibnamefont {Wang}}, \bibinfo
  {author} {\bibfnamefont {J.-d.}\ \bibnamefont {Zhang}}, \bibinfo {author}
  {\bibfnamefont {X.-D.}\ \bibnamefont {Li}}, \bibinfo {author} {\bibfnamefont
  {M.}~\bibnamefont {Hendry}}, \ and\ \bibinfo {author} {\bibfnamefont
  {J.}~\bibnamefont {Mei}},\ }\href {\doibase 10.1103/PhysRevResearch.4.013247}
  {\bibfield  {journal} {\bibinfo  {journal} {Phys. Rev. Res.}\ }\textbf
  {\bibinfo {volume} {4}},\ \bibinfo {pages} {013247} (\bibinfo {year}
  {2022})},\ \Eprint {http://arxiv.org/abs/2104.11956} {arXiv:2104.11956
  [astro-ph.CO]} \BibitemShut {NoStop}%
\bibitem [{\citenamefont {Auclair}\ \emph {et~al.}(2023)\citenamefont {Auclair}
  \emph {et~al.}}]{LISACosmologyWorkingGroup:2022jok}%
  \BibitemOpen
  \bibfield  {author} {\bibinfo {author} {\bibfnamefont {P.}~\bibnamefont
  {Auclair}} \emph {et~al.} (\bibinfo {collaboration} {LISA Cosmology Working
  Group}),\ }\href {\doibase 10.1007/s41114-023-00045-2} {\bibfield  {journal}
  {\bibinfo  {journal} {Living Rev. Rel.}\ }\textbf {\bibinfo {volume} {26}},\
  \bibinfo {pages} {5} (\bibinfo {year} {2023})},\ \Eprint
  {http://arxiv.org/abs/2204.05434} {arXiv:2204.05434 [astro-ph.CO]}
  \BibitemShut {NoStop}%
\bibitem [{\citenamefont {Caldwell}\ \emph {et~al.}(2019)\citenamefont
  {Caldwell} \emph {et~al.}}]{Caldwell:2019vru}%
  \BibitemOpen
  \bibfield  {author} {\bibinfo {author} {\bibfnamefont {R.}~\bibnamefont
  {Caldwell}} \emph {et~al.},\ }\href@noop {} {\  (\bibinfo {year} {2019})},\
  \Eprint {http://arxiv.org/abs/1903.04657} {arXiv:1903.04657 [astro-ph.CO]}
  \BibitemShut {NoStop}%
\bibitem [{\citenamefont {Shi}\ \emph {et~al.}(2019)\citenamefont {Shi},
  \citenamefont {Bao}, \citenamefont {Wang}, \citenamefont {Zhang},
  \citenamefont {Hu}, \citenamefont {Sesana}, \citenamefont {Barausse},
  \citenamefont {Mei},\ and\ \citenamefont {Luo}}]{Shi:2019hqa}%
  \BibitemOpen
  \bibfield  {author} {\bibinfo {author} {\bibfnamefont {C.}~\bibnamefont
  {Shi}}, \bibinfo {author} {\bibfnamefont {J.}~\bibnamefont {Bao}}, \bibinfo
  {author} {\bibfnamefont {H.}~\bibnamefont {Wang}}, \bibinfo {author}
  {\bibfnamefont {J.-d.}\ \bibnamefont {Zhang}}, \bibinfo {author}
  {\bibfnamefont {Y.}~\bibnamefont {Hu}}, \bibinfo {author} {\bibfnamefont
  {A.}~\bibnamefont {Sesana}}, \bibinfo {author} {\bibfnamefont
  {E.}~\bibnamefont {Barausse}}, \bibinfo {author} {\bibfnamefont
  {J.}~\bibnamefont {Mei}}, \ and\ \bibinfo {author} {\bibfnamefont
  {J.}~\bibnamefont {Luo}},\ }\href {\doibase 10.1103/PhysRevD.100.044036}
  {\bibfield  {journal} {\bibinfo  {journal} {Phys. Rev. D}\ }\textbf {\bibinfo
  {volume} {100}},\ \bibinfo {pages} {044036} (\bibinfo {year} {2019})},\
  \Eprint {http://arxiv.org/abs/1902.08922} {arXiv:1902.08922 [gr-qc]}
  \BibitemShut {NoStop}%
\bibitem [{\citenamefont {Shi}\ \emph {et~al.}(2023)\citenamefont {Shi},
  \citenamefont {Ji}, \citenamefont {Zhang},\ and\ \citenamefont
  {Mei}}]{Shi:2022qno}%
  \BibitemOpen
  \bibfield  {author} {\bibinfo {author} {\bibfnamefont {C.}~\bibnamefont
  {Shi}}, \bibinfo {author} {\bibfnamefont {M.}~\bibnamefont {Ji}}, \bibinfo
  {author} {\bibfnamefont {J.-d.}\ \bibnamefont {Zhang}}, \ and\ \bibinfo
  {author} {\bibfnamefont {J.}~\bibnamefont {Mei}},\ }\href {\doibase
  10.1103/PhysRevD.108.024030} {\bibfield  {journal} {\bibinfo  {journal}
  {Phys. Rev. D}\ }\textbf {\bibinfo {volume} {108}},\ \bibinfo {pages}
  {024030} (\bibinfo {year} {2023})},\ \Eprint
  {http://arxiv.org/abs/2210.13006} {arXiv:2210.13006 [gr-qc]} \BibitemShut
  {NoStop}%
\bibitem [{\citenamefont {Zi}\ \emph {et~al.}(2021)\citenamefont {Zi},
  \citenamefont {Zhang}, \citenamefont {Fan}, \citenamefont {Zhang},
  \citenamefont {Hu}, \citenamefont {Shi},\ and\ \citenamefont
  {Mei}}]{Zi:2021pdp}%
  \BibitemOpen
  \bibfield  {author} {\bibinfo {author} {\bibfnamefont {T.-G.}\ \bibnamefont
  {Zi}}, \bibinfo {author} {\bibfnamefont {J.-D.}\ \bibnamefont {Zhang}},
  \bibinfo {author} {\bibfnamefont {H.-M.}\ \bibnamefont {Fan}}, \bibinfo
  {author} {\bibfnamefont {X.-T.}\ \bibnamefont {Zhang}}, \bibinfo {author}
  {\bibfnamefont {Y.-M.}\ \bibnamefont {Hu}}, \bibinfo {author} {\bibfnamefont
  {C.}~\bibnamefont {Shi}}, \ and\ \bibinfo {author} {\bibfnamefont
  {J.}~\bibnamefont {Mei}},\ }\href@noop {} {\  (\bibinfo {year} {2021})},\
  \Eprint {http://arxiv.org/abs/2104.06047} {arXiv:2104.06047 [gr-qc]}
  \BibitemShut {NoStop}%
\bibitem [{\citenamefont {Xie}\ \emph {et~al.}(2022)\citenamefont {Xie},
  \citenamefont {Zhang}, \citenamefont {Huang}, \citenamefont {Hu},\ and\
  \citenamefont {Mei}}]{Xie:2022wkx}%
  \BibitemOpen
  \bibfield  {author} {\bibinfo {author} {\bibfnamefont {N.}~\bibnamefont
  {Xie}}, \bibinfo {author} {\bibfnamefont {J.-d.}\ \bibnamefont {Zhang}},
  \bibinfo {author} {\bibfnamefont {S.-J.}\ \bibnamefont {Huang}}, \bibinfo
  {author} {\bibfnamefont {Y.-M.}\ \bibnamefont {Hu}}, \ and\ \bibinfo {author}
  {\bibfnamefont {J.}~\bibnamefont {Mei}},\ }\href {\doibase
  10.1103/PhysRevD.106.124017} {\bibfield  {journal} {\bibinfo  {journal}
  {Phys. Rev. D}\ }\textbf {\bibinfo {volume} {106}},\ \bibinfo {pages}
  {124017} (\bibinfo {year} {2022})},\ \Eprint
  {http://arxiv.org/abs/2208.10831} {arXiv:2208.10831 [gr-qc]} \BibitemShut
  {NoStop}%
\bibitem [{\citenamefont {Rahman}\ \emph {et~al.}(2023)\citenamefont {Rahman},
  \citenamefont {Kumar},\ and\ \citenamefont {Bhattacharyya}}]{Rahman:2022fay}%
  \BibitemOpen
  \bibfield  {author} {\bibinfo {author} {\bibfnamefont {M.}~\bibnamefont
  {Rahman}}, \bibinfo {author} {\bibfnamefont {S.}~\bibnamefont {Kumar}}, \
  and\ \bibinfo {author} {\bibfnamefont {A.}~\bibnamefont {Bhattacharyya}},\
  }\href {\doibase 10.1088/1475-7516/2023/01/046} {\bibfield  {journal}
  {\bibinfo  {journal} {JCAP}\ }\textbf {\bibinfo {volume} {01}},\ \bibinfo
  {pages} {046} (\bibinfo {year} {2023})},\ \Eprint
  {http://arxiv.org/abs/2212.01404} {arXiv:2212.01404 [gr-qc]} \BibitemShut
  {NoStop}%
\bibitem [{\citenamefont {Barack}\ and\ \citenamefont
  {Cutler}(2004)}]{Barack:2003fp}%
  \BibitemOpen
  \bibfield  {author} {\bibinfo {author} {\bibfnamefont {L.}~\bibnamefont
  {Barack}}\ and\ \bibinfo {author} {\bibfnamefont {C.}~\bibnamefont
  {Cutler}},\ }\href {\doibase 10.1103/PhysRevD.69.082005} {\bibfield
  {journal} {\bibinfo  {journal} {Phys. Rev. D}\ }\textbf {\bibinfo {volume}
  {69}},\ \bibinfo {pages} {082005} (\bibinfo {year} {2004})},\ \Eprint
  {http://arxiv.org/abs/gr-qc/0310125} {arXiv:gr-qc/0310125} \BibitemShut
  {NoStop}%
\bibitem [{\citenamefont {Amaro-Seoane}\ \emph {et~al.}(2007)\citenamefont
  {Amaro-Seoane}, \citenamefont {Gair}, \citenamefont {Freitag}, \citenamefont
  {Coleman~Miller}, \citenamefont {Mandel}, \citenamefont {Cutler},\ and\
  \citenamefont {Babak}}]{Amaro-Seoane:2007osp}%
  \BibitemOpen
  \bibfield  {author} {\bibinfo {author} {\bibfnamefont {P.}~\bibnamefont
  {Amaro-Seoane}}, \bibinfo {author} {\bibfnamefont {J.~R.}\ \bibnamefont
  {Gair}}, \bibinfo {author} {\bibfnamefont {M.}~\bibnamefont {Freitag}},
  \bibinfo {author} {\bibfnamefont {M.}~\bibnamefont {Coleman~Miller}},
  \bibinfo {author} {\bibfnamefont {I.}~\bibnamefont {Mandel}}, \bibinfo
  {author} {\bibfnamefont {C.~J.}\ \bibnamefont {Cutler}}, \ and\ \bibinfo
  {author} {\bibfnamefont {S.}~\bibnamefont {Babak}},\ }\href {\doibase
  10.1088/0264-9381/24/17/R01} {\bibfield  {journal} {\bibinfo  {journal}
  {Class. Quant. Grav.}\ }\textbf {\bibinfo {volume} {24}},\ \bibinfo {pages}
  {R113} (\bibinfo {year} {2007})},\ \Eprint
  {http://arxiv.org/abs/astro-ph/0703495} {arXiv:astro-ph/0703495} \BibitemShut
  {NoStop}%
\bibitem [{\citenamefont {Chowdhuri}\ \emph {et~al.}(2024)\citenamefont
  {Chowdhuri}, \citenamefont {Bhattacharyya},\ and\ \citenamefont
  {Kumar}}]{AbhishekChowdhuri:2023wdf}%
  \BibitemOpen
  \bibfield  {author} {\bibinfo {author} {\bibfnamefont {A.}~\bibnamefont
  {Chowdhuri}}, \bibinfo {author} {\bibfnamefont {A.}~\bibnamefont
  {Bhattacharyya}}, \ and\ \bibinfo {author} {\bibfnamefont {S.}~\bibnamefont
  {Kumar}},\ }\href {\doibase 10.1088/1475-7516/2024/04/001} {\bibfield
  {journal} {\bibinfo  {journal} {JCAP}\ }\textbf {\bibinfo {volume} {04}},\
  \bibinfo {pages} {001} (\bibinfo {year} {2024})},\ \Eprint
  {http://arxiv.org/abs/2311.05983} {arXiv:2311.05983 [gr-qc]} \BibitemShut
  {NoStop}%
\bibitem [{\citenamefont {Kanti}\ \emph {et~al.}(1996)\citenamefont {Kanti},
  \citenamefont {Mavromatos}, \citenamefont {Rizos}, \citenamefont {Tamvakis},\
  and\ \citenamefont {Winstanley}}]{Kanti:1995vq}%
  \BibitemOpen
  \bibfield  {author} {\bibinfo {author} {\bibfnamefont {P.}~\bibnamefont
  {Kanti}}, \bibinfo {author} {\bibfnamefont {N.~E.}\ \bibnamefont
  {Mavromatos}}, \bibinfo {author} {\bibfnamefont {J.}~\bibnamefont {Rizos}},
  \bibinfo {author} {\bibfnamefont {K.}~\bibnamefont {Tamvakis}}, \ and\
  \bibinfo {author} {\bibfnamefont {E.}~\bibnamefont {Winstanley}},\ }\href
  {\doibase 10.1103/PhysRevD.54.5049} {\bibfield  {journal} {\bibinfo
  {journal} {Phys. Rev. D}\ }\textbf {\bibinfo {volume} {54}},\ \bibinfo
  {pages} {5049} (\bibinfo {year} {1996})},\ \Eprint
  {http://arxiv.org/abs/hep-th/9511071} {arXiv:hep-th/9511071} \BibitemShut
  {NoStop}%
\bibitem [{\citenamefont {Torii}\ \emph {et~al.}(1997)\citenamefont {Torii},
  \citenamefont {Yajima},\ and\ \citenamefont {Maeda}}]{Torii:1996yi}%
  \BibitemOpen
  \bibfield  {author} {\bibinfo {author} {\bibfnamefont {T.}~\bibnamefont
  {Torii}}, \bibinfo {author} {\bibfnamefont {H.}~\bibnamefont {Yajima}}, \
  and\ \bibinfo {author} {\bibfnamefont {K.-i.}\ \bibnamefont {Maeda}},\ }\href
  {\doibase 10.1103/PhysRevD.55.739} {\bibfield  {journal} {\bibinfo  {journal}
  {Phys. Rev. D}\ }\textbf {\bibinfo {volume} {55}},\ \bibinfo {pages} {739}
  (\bibinfo {year} {1997})},\ \Eprint {http://arxiv.org/abs/gr-qc/9606034}
  {arXiv:gr-qc/9606034} \BibitemShut {NoStop}%
\bibitem [{\citenamefont {Kanti}\ \emph {et~al.}(1998)\citenamefont {Kanti},
  \citenamefont {Mavromatos}, \citenamefont {Rizos}, \citenamefont {Tamvakis},\
  and\ \citenamefont {Winstanley}}]{Kanti:1997br}%
  \BibitemOpen
  \bibfield  {author} {\bibinfo {author} {\bibfnamefont {P.}~\bibnamefont
  {Kanti}}, \bibinfo {author} {\bibfnamefont {N.~E.}\ \bibnamefont
  {Mavromatos}}, \bibinfo {author} {\bibfnamefont {J.}~\bibnamefont {Rizos}},
  \bibinfo {author} {\bibfnamefont {K.}~\bibnamefont {Tamvakis}}, \ and\
  \bibinfo {author} {\bibfnamefont {E.}~\bibnamefont {Winstanley}},\ }\href
  {\doibase 10.1103/PhysRevD.57.6255} {\bibfield  {journal} {\bibinfo
  {journal} {Phys. Rev. D}\ }\textbf {\bibinfo {volume} {57}},\ \bibinfo
  {pages} {6255} (\bibinfo {year} {1998})},\ \Eprint
  {http://arxiv.org/abs/hep-th/9703192} {arXiv:hep-th/9703192} \BibitemShut
  {NoStop}%
\bibitem [{\citenamefont {Nojiri}\ and\ \citenamefont
  {Odintsov}(2011)}]{Nojiri:2010wj}%
  \BibitemOpen
  \bibfield  {author} {\bibinfo {author} {\bibfnamefont {S.}~\bibnamefont
  {Nojiri}}\ and\ \bibinfo {author} {\bibfnamefont {S.~D.}\ \bibnamefont
  {Odintsov}},\ }\href {\doibase 10.1016/j.physrep.2011.04.001} {\bibfield
  {journal} {\bibinfo  {journal} {Phys. Rept.}\ }\textbf {\bibinfo {volume}
  {505}},\ \bibinfo {pages} {59} (\bibinfo {year} {2011})},\ \Eprint
  {http://arxiv.org/abs/1011.0544} {arXiv:1011.0544 [gr-qc]} \BibitemShut
  {NoStop}%
\bibitem [{\citenamefont {Pomazanov}\ \emph {et~al.}(2003)\citenamefont
  {Pomazanov}, \citenamefont {Kolubasova},\ and\ \citenamefont
  {Alexeyev}}]{Pomazanov:2003wq}%
  \BibitemOpen
  \bibfield  {author} {\bibinfo {author} {\bibfnamefont {M.}~\bibnamefont
  {Pomazanov}}, \bibinfo {author} {\bibfnamefont {V.}~\bibnamefont
  {Kolubasova}}, \ and\ \bibinfo {author} {\bibfnamefont {S.}~\bibnamefont
  {Alexeyev}},\ }\href@noop {} {\  (\bibinfo {year} {2003})},\ \Eprint
  {http://arxiv.org/abs/gr-qc/0301029} {arXiv:gr-qc/0301029} \BibitemShut
  {NoStop}%
\bibitem [{\citenamefont {Pani}\ and\ \citenamefont
  {Cardoso}(2009)}]{Pani:2009wy}%
  \BibitemOpen
  \bibfield  {author} {\bibinfo {author} {\bibfnamefont {P.}~\bibnamefont
  {Pani}}\ and\ \bibinfo {author} {\bibfnamefont {V.}~\bibnamefont {Cardoso}},\
  }\href {\doibase 10.1103/PhysRevD.79.084031} {\bibfield  {journal} {\bibinfo
  {journal} {Phys. Rev. D}\ }\textbf {\bibinfo {volume} {79}},\ \bibinfo
  {pages} {084031} (\bibinfo {year} {2009})},\ \Eprint
  {http://arxiv.org/abs/0902.1569} {arXiv:0902.1569 [gr-qc]} \BibitemShut
  {NoStop}%
\bibitem [{\citenamefont {Amendola}\ \emph {et~al.}(2007)\citenamefont
  {Amendola}, \citenamefont {Charmousis},\ and\ \citenamefont
  {Davis}}]{Amendola:2007ni}%
  \BibitemOpen
  \bibfield  {author} {\bibinfo {author} {\bibfnamefont {L.}~\bibnamefont
  {Amendola}}, \bibinfo {author} {\bibfnamefont {C.}~\bibnamefont
  {Charmousis}}, \ and\ \bibinfo {author} {\bibfnamefont {S.~C.}\ \bibnamefont
  {Davis}},\ }\href {\doibase 10.1088/1475-7516/2007/10/004} {\bibfield
  {journal} {\bibinfo  {journal} {JCAP}\ }\textbf {\bibinfo {volume} {10}},\
  \bibinfo {pages} {004} (\bibinfo {year} {2007})},\ \Eprint
  {http://arxiv.org/abs/0704.0175} {arXiv:0704.0175 [astro-ph]} \BibitemShut
  {NoStop}%
\bibitem [{\citenamefont {Yagi}(2012)}]{Yagi:2012gp}%
  \BibitemOpen
  \bibfield  {author} {\bibinfo {author} {\bibfnamefont {K.}~\bibnamefont
  {Yagi}},\ }\href {\doibase 10.1103/PhysRevD.86.081504} {\bibfield  {journal}
  {\bibinfo  {journal} {Phys. Rev. D}\ }\textbf {\bibinfo {volume} {86}},\
  \bibinfo {pages} {081504} (\bibinfo {year} {2012})},\ \Eprint
  {http://arxiv.org/abs/1204.4524} {arXiv:1204.4524 [gr-qc]} \BibitemShut
  {NoStop}%
\bibitem [{\citenamefont {Nair}\ \emph {et~al.}(2019)\citenamefont {Nair},
  \citenamefont {Perkins}, \citenamefont {Silva},\ and\ \citenamefont
  {Yunes}}]{Nair:2019iur}%
  \BibitemOpen
  \bibfield  {author} {\bibinfo {author} {\bibfnamefont {R.}~\bibnamefont
  {Nair}}, \bibinfo {author} {\bibfnamefont {S.}~\bibnamefont {Perkins}},
  \bibinfo {author} {\bibfnamefont {H.~O.}\ \bibnamefont {Silva}}, \ and\
  \bibinfo {author} {\bibfnamefont {N.}~\bibnamefont {Yunes}},\ }\href
  {\doibase 10.1103/PhysRevLett.123.191101} {\bibfield  {journal} {\bibinfo
  {journal} {Phys. Rev. Lett.}\ }\textbf {\bibinfo {volume} {123}},\ \bibinfo
  {pages} {191101} (\bibinfo {year} {2019})},\ \Eprint
  {http://arxiv.org/abs/1905.00870} {arXiv:1905.00870 [gr-qc]} \BibitemShut
  {NoStop}%
\bibitem [{\citenamefont {Yamada}\ \emph {et~al.}(2019)\citenamefont {Yamada},
  \citenamefont {Narikawa},\ and\ \citenamefont {Tanaka}}]{Yamada:2019zrb}%
  \BibitemOpen
  \bibfield  {author} {\bibinfo {author} {\bibfnamefont {K.}~\bibnamefont
  {Yamada}}, \bibinfo {author} {\bibfnamefont {T.}~\bibnamefont {Narikawa}}, \
  and\ \bibinfo {author} {\bibfnamefont {T.}~\bibnamefont {Tanaka}},\ }\href
  {\doibase 10.1093/ptep/ptz103} {\bibfield  {journal} {\bibinfo  {journal}
  {PTEP}\ }\textbf {\bibinfo {volume} {2019}},\ \bibinfo {pages} {103E01}
  (\bibinfo {year} {2019})},\ \Eprint {http://arxiv.org/abs/1905.11859}
  {arXiv:1905.11859 [gr-qc]} \BibitemShut {NoStop}%
\bibitem [{\citenamefont {Tahura}\ \emph {et~al.}(2019)\citenamefont {Tahura},
  \citenamefont {Yagi},\ and\ \citenamefont {Carson}}]{Tahura:2019dgr}%
  \BibitemOpen
  \bibfield  {author} {\bibinfo {author} {\bibfnamefont {S.}~\bibnamefont
  {Tahura}}, \bibinfo {author} {\bibfnamefont {K.}~\bibnamefont {Yagi}}, \ and\
  \bibinfo {author} {\bibfnamefont {Z.}~\bibnamefont {Carson}},\ }\href
  {\doibase 10.1103/PhysRevD.100.104001} {\bibfield  {journal} {\bibinfo
  {journal} {Phys. Rev. D}\ }\textbf {\bibinfo {volume} {100}},\ \bibinfo
  {pages} {104001} (\bibinfo {year} {2019})},\ \Eprint
  {http://arxiv.org/abs/1907.10059} {arXiv:1907.10059 [gr-qc]} \BibitemShut
  {NoStop}%
\bibitem [{\citenamefont {Carson}\ and\ \citenamefont
  {Yagi}(2020{\natexlab{a}})}]{Carson:2020ter}%
  \BibitemOpen
  \bibfield  {author} {\bibinfo {author} {\bibfnamefont {Z.}~\bibnamefont
  {Carson}}\ and\ \bibinfo {author} {\bibfnamefont {K.}~\bibnamefont {Yagi}},\
  }\href {\doibase 10.1103/PhysRevD.101.104030} {\bibfield  {journal} {\bibinfo
   {journal} {Phys. Rev. D}\ }\textbf {\bibinfo {volume} {101}},\ \bibinfo
  {pages} {104030} (\bibinfo {year} {2020}{\natexlab{a}})},\ \Eprint
  {http://arxiv.org/abs/2003.00286} {arXiv:2003.00286 [gr-qc]} \BibitemShut
  {NoStop}%
\bibitem [{\citenamefont {Okounkova}(2020)}]{Okounkova:2020rqw}%
  \BibitemOpen
  \bibfield  {author} {\bibinfo {author} {\bibfnamefont {M.}~\bibnamefont
  {Okounkova}},\ }\href {\doibase 10.1103/PhysRevD.102.084046} {\bibfield
  {journal} {\bibinfo  {journal} {Phys. Rev. D}\ }\textbf {\bibinfo {volume}
  {102}},\ \bibinfo {pages} {084046} (\bibinfo {year} {2020})},\ \Eprint
  {http://arxiv.org/abs/2001.03571} {arXiv:2001.03571 [gr-qc]} \BibitemShut
  {NoStop}%
\bibitem [{\citenamefont {Perkins}\ \emph {et~al.}(2021)\citenamefont
  {Perkins}, \citenamefont {Nair}, \citenamefont {Silva},\ and\ \citenamefont
  {Yunes}}]{Perkins:2021mhb}%
  \BibitemOpen
  \bibfield  {author} {\bibinfo {author} {\bibfnamefont {S.~E.}\ \bibnamefont
  {Perkins}}, \bibinfo {author} {\bibfnamefont {R.}~\bibnamefont {Nair}},
  \bibinfo {author} {\bibfnamefont {H.~O.}\ \bibnamefont {Silva}}, \ and\
  \bibinfo {author} {\bibfnamefont {N.}~\bibnamefont {Yunes}},\ }\href
  {\doibase 10.1103/PhysRevD.104.024060} {\bibfield  {journal} {\bibinfo
  {journal} {Phys. Rev. D}\ }\textbf {\bibinfo {volume} {104}},\ \bibinfo
  {pages} {024060} (\bibinfo {year} {2021})},\ \Eprint
  {http://arxiv.org/abs/2104.11189} {arXiv:2104.11189 [gr-qc]} \BibitemShut
  {NoStop}%
\bibitem [{\citenamefont {Wang}\ \emph {et~al.}(2021)\citenamefont {Wang},
  \citenamefont {Tang}, \citenamefont {Li}, \citenamefont {Han},\ and\
  \citenamefont {Fan}}]{Wang:2021jfc}%
  \BibitemOpen
  \bibfield  {author} {\bibinfo {author} {\bibfnamefont {H.-T.}\ \bibnamefont
  {Wang}}, \bibinfo {author} {\bibfnamefont {S.-P.}\ \bibnamefont {Tang}},
  \bibinfo {author} {\bibfnamefont {P.-C.}\ \bibnamefont {Li}}, \bibinfo
  {author} {\bibfnamefont {M.-Z.}\ \bibnamefont {Han}}, \ and\ \bibinfo
  {author} {\bibfnamefont {Y.-Z.}\ \bibnamefont {Fan}},\ }\href {\doibase
  10.1103/PhysRevD.104.024015} {\bibfield  {journal} {\bibinfo  {journal}
  {Phys. Rev. D}\ }\textbf {\bibinfo {volume} {104}},\ \bibinfo {pages}
  {024015} (\bibinfo {year} {2021})}\BibitemShut {NoStop}%
\bibitem [{\citenamefont {Wang}\ \emph {et~al.}(2023)\citenamefont {Wang},
  \citenamefont {Shi}, \citenamefont {Zhang}, \citenamefont {hu},\ and\
  \citenamefont {Mei}}]{Wang:2023wgv}%
  \BibitemOpen
  \bibfield  {author} {\bibinfo {author} {\bibfnamefont {B.}~\bibnamefont
  {Wang}}, \bibinfo {author} {\bibfnamefont {C.}~\bibnamefont {Shi}}, \bibinfo
  {author} {\bibfnamefont {J.-d.}\ \bibnamefont {Zhang}}, \bibinfo {author}
  {\bibfnamefont {Y.-M.}\ \bibnamefont {hu}}, \ and\ \bibinfo {author}
  {\bibfnamefont {J.}~\bibnamefont {Mei}},\ }\href {\doibase
  10.1103/PhysRevD.108.044061} {\bibfield  {journal} {\bibinfo  {journal}
  {Phys. Rev. D}\ }\textbf {\bibinfo {volume} {108}},\ \bibinfo {pages}
  {044061} (\bibinfo {year} {2023})},\ \Eprint
  {http://arxiv.org/abs/2302.10112} {arXiv:2302.10112 [gr-qc]} \BibitemShut
  {NoStop}%
\bibitem [{\citenamefont {Odintsov}\ \emph {et~al.}(2020)\citenamefont
  {Odintsov}, \citenamefont {Oikonomou},\ and\ \citenamefont
  {Fronimos}}]{Odintsov:2020xji}%
  \BibitemOpen
  \bibfield  {author} {\bibinfo {author} {\bibfnamefont {S.~D.}\ \bibnamefont
  {Odintsov}}, \bibinfo {author} {\bibfnamefont {V.~K.}\ \bibnamefont
  {Oikonomou}}, \ and\ \bibinfo {author} {\bibfnamefont {F.~P.}\ \bibnamefont
  {Fronimos}},\ }\href {\doibase 10.1016/j.aop.2020.168250} {\bibfield
  {journal} {\bibinfo  {journal} {Annals Phys.}\ }\textbf {\bibinfo {volume}
  {420}},\ \bibinfo {pages} {168250} (\bibinfo {year} {2020})},\ \Eprint
  {http://arxiv.org/abs/2007.02309} {arXiv:2007.02309 [gr-qc]} \BibitemShut
  {NoStop}%
\bibitem [{\citenamefont {Luo}\ \emph {et~al.}(2024)\citenamefont {Luo},
  \citenamefont {Liu},\ and\ \citenamefont {Guo}}]{Luo:2024vls}%
  \BibitemOpen
  \bibfield  {author} {\bibinfo {author} {\bibfnamefont {W.}~\bibnamefont
  {Luo}}, \bibinfo {author} {\bibfnamefont {C.}~\bibnamefont {Liu}}, \ and\
  \bibinfo {author} {\bibfnamefont {Z.-K.}\ \bibnamefont {Guo}},\ }\href@noop
  {} {\  (\bibinfo {year} {2024})},\ \Eprint {http://arxiv.org/abs/2401.03669}
  {arXiv:2401.03669 [gr-qc]} \BibitemShut {NoStop}%
\bibitem [{\citenamefont {Shao}\ \emph {et~al.}(2023)\citenamefont {Shao},
  \citenamefont {Hu},\ and\ \citenamefont {Shao}}]{Shao:2023yjx}%
  \BibitemOpen
  \bibfield  {author} {\bibinfo {author} {\bibfnamefont {C.-Y.}\ \bibnamefont
  {Shao}}, \bibinfo {author} {\bibfnamefont {Y.}~\bibnamefont {Hu}}, \ and\
  \bibinfo {author} {\bibfnamefont {C.-G.}\ \bibnamefont {Shao}},\ }\href
  {\doibase 10.1088/1674-1137/ace522} {\bibfield  {journal} {\bibinfo
  {journal} {Chin. Phys. C}\ }\textbf {\bibinfo {volume} {47}},\ \bibinfo
  {pages} {105101} (\bibinfo {year} {2023})},\ \Eprint
  {http://arxiv.org/abs/2307.02084} {arXiv:2307.02084 [gr-qc]} \BibitemShut
  {NoStop}%
\bibitem [{\citenamefont {Carson}\ and\ \citenamefont
  {Yagi}(2020{\natexlab{b}})}]{Carson:2020cqb}%
  \BibitemOpen
  \bibfield  {author} {\bibinfo {author} {\bibfnamefont {Z.}~\bibnamefont
  {Carson}}\ and\ \bibinfo {author} {\bibfnamefont {K.}~\bibnamefont {Yagi}},\
  }\href {\doibase 10.1088/1361-6382/aba221} {\bibfield  {journal} {\bibinfo
  {journal} {Class. Quant. Grav.}\ }\textbf {\bibinfo {volume} {37}},\ \bibinfo
  {pages} {215007} (\bibinfo {year} {2020}{\natexlab{b}})},\ \Eprint
  {http://arxiv.org/abs/2002.08559} {arXiv:2002.08559 [gr-qc]} \BibitemShut
  {NoStop}%
\bibitem [{\citenamefont {Poisson}\ \emph {et~al.}(2011)\citenamefont
  {Poisson}, \citenamefont {Pound},\ and\ \citenamefont
  {Vega}}]{Poisson:2011nh}%
  \BibitemOpen
  \bibfield  {author} {\bibinfo {author} {\bibfnamefont {E.}~\bibnamefont
  {Poisson}}, \bibinfo {author} {\bibfnamefont {A.}~\bibnamefont {Pound}}, \
  and\ \bibinfo {author} {\bibfnamefont {I.}~\bibnamefont {Vega}},\ }\href
  {\doibase 10.12942/lrr-2011-7} {\bibfield  {journal} {\bibinfo  {journal}
  {Living Rev. Rel.}\ }\textbf {\bibinfo {volume} {14}},\ \bibinfo {pages} {7}
  (\bibinfo {year} {2011})},\ \Eprint {http://arxiv.org/abs/1102.0529}
  {arXiv:1102.0529 [gr-qc]} \BibitemShut {NoStop}%
\bibitem [{\citenamefont {Osburn}\ \emph {et~al.}(2016)\citenamefont {Osburn},
  \citenamefont {Warburton},\ and\ \citenamefont {Evans}}]{Osburn:2015duj}%
  \BibitemOpen
  \bibfield  {author} {\bibinfo {author} {\bibfnamefont {T.}~\bibnamefont
  {Osburn}}, \bibinfo {author} {\bibfnamefont {N.}~\bibnamefont {Warburton}}, \
  and\ \bibinfo {author} {\bibfnamefont {C.~R.}\ \bibnamefont {Evans}},\ }\href
  {\doibase 10.1103/PhysRevD.93.064024} {\bibfield  {journal} {\bibinfo
  {journal} {Phys. Rev. D}\ }\textbf {\bibinfo {volume} {93}},\ \bibinfo
  {pages} {064024} (\bibinfo {year} {2016})},\ \Eprint
  {http://arxiv.org/abs/1511.01498} {arXiv:1511.01498 [gr-qc]} \BibitemShut
  {NoStop}%
\bibitem [{\citenamefont {Warburton}\ \emph {et~al.}(2017)\citenamefont
  {Warburton}, \citenamefont {Osburn},\ and\ \citenamefont
  {Evans}}]{Warburton:2017sxk}%
  \BibitemOpen
  \bibfield  {author} {\bibinfo {author} {\bibfnamefont {N.}~\bibnamefont
  {Warburton}}, \bibinfo {author} {\bibfnamefont {T.}~\bibnamefont {Osburn}}, \
  and\ \bibinfo {author} {\bibfnamefont {C.~R.}\ \bibnamefont {Evans}},\ }\href
  {\doibase 10.1103/PhysRevD.96.084057} {\bibfield  {journal} {\bibinfo
  {journal} {Phys. Rev. D}\ }\textbf {\bibinfo {volume} {96}},\ \bibinfo
  {pages} {084057} (\bibinfo {year} {2017})},\ \Eprint
  {http://arxiv.org/abs/1708.03720} {arXiv:1708.03720 [gr-qc]} \BibitemShut
  {NoStop}%
\bibitem [{\citenamefont {van~de Meent}(2018)}]{vandeMeent:2017bcc}%
  \BibitemOpen
  \bibfield  {author} {\bibinfo {author} {\bibfnamefont {M.}~\bibnamefont
  {van~de Meent}},\ }\href {\doibase 10.1103/PhysRevD.97.104033} {\bibfield
  {journal} {\bibinfo  {journal} {Phys. Rev. D}\ }\textbf {\bibinfo {volume}
  {97}},\ \bibinfo {pages} {104033} (\bibinfo {year} {2018})},\ \Eprint
  {http://arxiv.org/abs/1711.09607} {arXiv:1711.09607 [gr-qc]} \BibitemShut
  {NoStop}%
\bibitem [{\citenamefont {Babak}\ \emph {et~al.}(2007)\citenamefont {Babak},
  \citenamefont {Fang}, \citenamefont {Gair}, \citenamefont {Glampedakis},\
  and\ \citenamefont {Hughes}}]{Babak:2006uv}%
  \BibitemOpen
  \bibfield  {author} {\bibinfo {author} {\bibfnamefont {S.}~\bibnamefont
  {Babak}}, \bibinfo {author} {\bibfnamefont {H.}~\bibnamefont {Fang}},
  \bibinfo {author} {\bibfnamefont {J.~R.}\ \bibnamefont {Gair}}, \bibinfo
  {author} {\bibfnamefont {K.}~\bibnamefont {Glampedakis}}, \ and\ \bibinfo
  {author} {\bibfnamefont {S.~A.}\ \bibnamefont {Hughes}},\ }\href {\doibase
  10.1103/PhysRevD.75.024005} {\bibfield  {journal} {\bibinfo  {journal} {Phys.
  Rev. D}\ }\textbf {\bibinfo {volume} {75}},\ \bibinfo {pages} {024005}
  (\bibinfo {year} {2007})},\ \bibinfo {note} {[Erratum: Phys.Rev.D 77, 04990
  (2008)]},\ \Eprint {http://arxiv.org/abs/gr-qc/0607007} {arXiv:gr-qc/0607007}
  \BibitemShut {NoStop}%
\bibitem [{\citenamefont {Chua}\ \emph {et~al.}(2017)\citenamefont {Chua},
  \citenamefont {Moore},\ and\ \citenamefont {Gair}}]{Chua:2017ujo}%
  \BibitemOpen
  \bibfield  {author} {\bibinfo {author} {\bibfnamefont {A.~J.~K.}\
  \bibnamefont {Chua}}, \bibinfo {author} {\bibfnamefont {C.~J.}\ \bibnamefont
  {Moore}}, \ and\ \bibinfo {author} {\bibfnamefont {J.~R.}\ \bibnamefont
  {Gair}},\ }\href {\doibase 10.1103/PhysRevD.96.044005} {\bibfield  {journal}
  {\bibinfo  {journal} {Phys. Rev. D}\ }\textbf {\bibinfo {volume} {96}},\
  \bibinfo {pages} {044005} (\bibinfo {year} {2017})},\ \Eprint
  {http://arxiv.org/abs/1705.04259} {arXiv:1705.04259 [gr-qc]} \BibitemShut
  {NoStop}%
\bibitem [{\citenamefont {Liu}\ and\ \citenamefont
  {Zhang}(2020)}]{Liu:2020ghq}%
  \BibitemOpen
  \bibfield  {author} {\bibinfo {author} {\bibfnamefont {M.}~\bibnamefont
  {Liu}}\ and\ \bibinfo {author} {\bibfnamefont {J.-d.}\ \bibnamefont
  {Zhang}},\ }\href@noop {} {\  (\bibinfo {year} {2020})},\ \Eprint
  {http://arxiv.org/abs/2008.11396} {arXiv:2008.11396 [gr-qc]} \BibitemShut
  {NoStop}%
\bibitem [{\citenamefont {Katz}\ \emph {et~al.}(2021)\citenamefont {Katz},
  \citenamefont {Chua}, \citenamefont {Speri}, \citenamefont {Warburton},\ and\
  \citenamefont {Hughes}}]{Katz:2021yft}%
  \BibitemOpen
  \bibfield  {author} {\bibinfo {author} {\bibfnamefont {M.~L.}\ \bibnamefont
  {Katz}}, \bibinfo {author} {\bibfnamefont {A.~J.~K.}\ \bibnamefont {Chua}},
  \bibinfo {author} {\bibfnamefont {L.}~\bibnamefont {Speri}}, \bibinfo
  {author} {\bibfnamefont {N.}~\bibnamefont {Warburton}}, \ and\ \bibinfo
  {author} {\bibfnamefont {S.~A.}\ \bibnamefont {Hughes}},\ }\href {\doibase
  10.1103/PhysRevD.104.064047} {\bibfield  {journal} {\bibinfo  {journal}
  {Phys. Rev. D}\ }\textbf {\bibinfo {volume} {104}},\ \bibinfo {pages}
  {064047} (\bibinfo {year} {2021})},\ \Eprint
  {http://arxiv.org/abs/2104.04582} {arXiv:2104.04582 [gr-qc]} \BibitemShut
  {NoStop}%
\bibitem [{\citenamefont {Gair}\ and\ \citenamefont
  {Glampedakis}(2006)}]{Gair:2005ih}%
  \BibitemOpen
  \bibfield  {author} {\bibinfo {author} {\bibfnamefont {J.~R.}\ \bibnamefont
  {Gair}}\ and\ \bibinfo {author} {\bibfnamefont {K.}~\bibnamefont
  {Glampedakis}},\ }\href {\doibase 10.1103/PhysRevD.73.064037} {\bibfield
  {journal} {\bibinfo  {journal} {Phys. Rev. D}\ }\textbf {\bibinfo {volume}
  {73}},\ \bibinfo {pages} {064037} (\bibinfo {year} {2006})},\ \Eprint
  {http://arxiv.org/abs/gr-qc/0510129} {arXiv:gr-qc/0510129} \BibitemShut
  {NoStop}%
\bibitem [{\citenamefont {Yagi}\ \emph {et~al.}(2012)\citenamefont {Yagi},
  \citenamefont {Stein}, \citenamefont {Yunes},\ and\ \citenamefont
  {Tanaka}}]{Yagi:2011xp}%
  \BibitemOpen
  \bibfield  {author} {\bibinfo {author} {\bibfnamefont {K.}~\bibnamefont
  {Yagi}}, \bibinfo {author} {\bibfnamefont {L.~C.}\ \bibnamefont {Stein}},
  \bibinfo {author} {\bibfnamefont {N.}~\bibnamefont {Yunes}}, \ and\ \bibinfo
  {author} {\bibfnamefont {T.}~\bibnamefont {Tanaka}},\ }\href {\doibase
  10.1103/PhysRevD.85.064022} {\bibfield  {journal} {\bibinfo  {journal} {Phys.
  Rev. D}\ }\textbf {\bibinfo {volume} {85}},\ \bibinfo {pages} {064022}
  (\bibinfo {year} {2012})},\ \bibinfo {note} {[Erratum: Phys.Rev.D 93, 029902
  (2016)]},\ \Eprint {http://arxiv.org/abs/1110.5950} {arXiv:1110.5950 [gr-qc]}
  \BibitemShut {NoStop}%
\bibitem [{\citenamefont {Yunes}\ and\ \citenamefont
  {Stein}(2011)}]{Yunes:2011we}%
  \BibitemOpen
  \bibfield  {author} {\bibinfo {author} {\bibfnamefont {N.}~\bibnamefont
  {Yunes}}\ and\ \bibinfo {author} {\bibfnamefont {L.~C.}\ \bibnamefont
  {Stein}},\ }\href {\doibase 10.1103/PhysRevD.83.104002} {\bibfield  {journal}
  {\bibinfo  {journal} {Phys. Rev. D}\ }\textbf {\bibinfo {volume} {83}},\
  \bibinfo {pages} {104002} (\bibinfo {year} {2011})},\ \Eprint
  {http://arxiv.org/abs/1101.2921} {arXiv:1101.2921 [gr-qc]} \BibitemShut
  {NoStop}%
\bibitem [{\citenamefont {Maselli}\ \emph {et~al.}(2020)\citenamefont
  {Maselli}, \citenamefont {Franchini}, \citenamefont {Gualtieri},\ and\
  \citenamefont {Sotiriou}}]{Maselli:2020zgv}%
  \BibitemOpen
  \bibfield  {author} {\bibinfo {author} {\bibfnamefont {A.}~\bibnamefont
  {Maselli}}, \bibinfo {author} {\bibfnamefont {N.}~\bibnamefont {Franchini}},
  \bibinfo {author} {\bibfnamefont {L.}~\bibnamefont {Gualtieri}}, \ and\
  \bibinfo {author} {\bibfnamefont {T.~P.}\ \bibnamefont {Sotiriou}},\ }\href
  {\doibase 10.1103/PhysRevLett.125.141101} {\bibfield  {journal} {\bibinfo
  {journal} {Phys. Rev. Lett.}\ }\textbf {\bibinfo {volume} {125}},\ \bibinfo
  {pages} {141101} (\bibinfo {year} {2020})},\ \Eprint
  {http://arxiv.org/abs/2004.11895} {arXiv:2004.11895 [gr-qc]} \BibitemShut
  {NoStop}%
\bibitem [{\citenamefont {Hughes}(2000)}]{Hughes:1999bq}%
  \BibitemOpen
  \bibfield  {author} {\bibinfo {author} {\bibfnamefont {S.~A.}\ \bibnamefont
  {Hughes}},\ }\href {\doibase 10.1103/PhysRevD.65.069902} {\bibfield
  {journal} {\bibinfo  {journal} {Phys. Rev. D}\ }\textbf {\bibinfo {volume}
  {61}},\ \bibinfo {pages} {084004} (\bibinfo {year} {2000})},\ \bibinfo {note}
  {[Erratum: Phys.Rev.D 63, 049902 (2001), Erratum: Phys.Rev.D 65, 069902
  (2002), Erratum: Phys.Rev.D 67, 089901 (2003), Erratum: Phys.Rev.D 78, 109902
  (2008), Erratum: Phys.Rev.D 90, 109904 (2014)]},\ \Eprint
  {http://arxiv.org/abs/gr-qc/9910091} {arXiv:gr-qc/9910091} \BibitemShut
  {NoStop}%
\bibitem [{\citenamefont {Hughes}(2001)}]{Hughes:2001jr}%
  \BibitemOpen
  \bibfield  {author} {\bibinfo {author} {\bibfnamefont {S.~A.}\ \bibnamefont
  {Hughes}},\ }\href {\doibase 10.1103/PhysRevD.64.064004} {\bibfield
  {journal} {\bibinfo  {journal} {Phys. Rev. D}\ }\textbf {\bibinfo {volume}
  {64}},\ \bibinfo {pages} {064004} (\bibinfo {year} {2001})},\ \bibinfo {note}
  {[Erratum: Phys.Rev.D 88, 109902 (2013)]},\ \Eprint
  {http://arxiv.org/abs/gr-qc/0104041} {arXiv:gr-qc/0104041} \BibitemShut
  {NoStop}%
\bibitem [{\citenamefont {Glampedakis}\ \emph {et~al.}(2002)\citenamefont
  {Glampedakis}, \citenamefont {Hughes},\ and\ \citenamefont
  {Kennefick}}]{Glampedakis:2002cb}%
  \BibitemOpen
  \bibfield  {author} {\bibinfo {author} {\bibfnamefont {K.}~\bibnamefont
  {Glampedakis}}, \bibinfo {author} {\bibfnamefont {S.~A.}\ \bibnamefont
  {Hughes}}, \ and\ \bibinfo {author} {\bibfnamefont {D.}~\bibnamefont
  {Kennefick}},\ }\href {\doibase 10.1103/PhysRevD.66.064005} {\bibfield
  {journal} {\bibinfo  {journal} {Phys. Rev. D}\ }\textbf {\bibinfo {volume}
  {66}},\ \bibinfo {pages} {064005} (\bibinfo {year} {2002})},\ \Eprint
  {http://arxiv.org/abs/gr-qc/0205033} {arXiv:gr-qc/0205033} \BibitemShut
  {NoStop}%
\bibitem [{\citenamefont {Peters}\ and\ \citenamefont
  {Mathews}(1963)}]{Peters:1963ux}%
  \BibitemOpen
  \bibfield  {author} {\bibinfo {author} {\bibfnamefont {P.~C.}\ \bibnamefont
  {Peters}}\ and\ \bibinfo {author} {\bibfnamefont {J.}~\bibnamefont
  {Mathews}},\ }\href {\doibase 10.1103/PhysRev.131.435} {\bibfield  {journal}
  {\bibinfo  {journal} {Phys. Rev.}\ }\textbf {\bibinfo {volume} {131}},\
  \bibinfo {pages} {435} (\bibinfo {year} {1963})}\BibitemShut {NoStop}%
\bibitem [{\citenamefont {Peters}(1964)}]{Peters:1964zz}%
  \BibitemOpen
  \bibfield  {author} {\bibinfo {author} {\bibfnamefont {P.~C.}\ \bibnamefont
  {Peters}},\ }\href {\doibase 10.1103/PhysRev.136.B1224} {\bibfield  {journal}
  {\bibinfo  {journal} {Phys. Rev.}\ }\textbf {\bibinfo {volume} {136}},\
  \bibinfo {pages} {B1224} (\bibinfo {year} {1964})}\BibitemShut {NoStop}%
\bibitem [{\citenamefont {Loutrel}\ \emph {et~al.}(2014)\citenamefont
  {Loutrel}, \citenamefont {Yunes},\ and\ \citenamefont
  {Pretorius}}]{Loutrel:2014vja}%
  \BibitemOpen
  \bibfield  {author} {\bibinfo {author} {\bibfnamefont {N.}~\bibnamefont
  {Loutrel}}, \bibinfo {author} {\bibfnamefont {N.}~\bibnamefont {Yunes}}, \
  and\ \bibinfo {author} {\bibfnamefont {F.}~\bibnamefont {Pretorius}},\ }\href
  {\doibase 10.1103/PhysRevD.90.104010} {\bibfield  {journal} {\bibinfo
  {journal} {Phys. Rev. D}\ }\textbf {\bibinfo {volume} {90}},\ \bibinfo
  {pages} {104010} (\bibinfo {year} {2014})},\ \Eprint
  {http://arxiv.org/abs/1404.0092} {arXiv:1404.0092 [gr-qc]} \BibitemShut
  {NoStop}%
\bibitem [{\citenamefont {Maggiore}(2000)}]{Maggiore:1999vm}%
  \BibitemOpen
  \bibfield  {author} {\bibinfo {author} {\bibfnamefont {M.}~\bibnamefont
  {Maggiore}},\ }\href {\doibase 10.1016/S0370-1573(99)00102-7} {\bibfield
  {journal} {\bibinfo  {journal} {Phys. Rept.}\ }\textbf {\bibinfo {volume}
  {331}},\ \bibinfo {pages} {283} (\bibinfo {year} {2000})},\ \Eprint
  {http://arxiv.org/abs/gr-qc/9909001} {arXiv:gr-qc/9909001} \BibitemShut
  {NoStop}%
\bibitem [{\citenamefont {Apostolatos}\ \emph {et~al.}(1994)\citenamefont
  {Apostolatos}, \citenamefont {Cutler}, \citenamefont {Sussman},\ and\
  \citenamefont {Thorne}}]{Apostolatos:1994mx}%
  \BibitemOpen
  \bibfield  {author} {\bibinfo {author} {\bibfnamefont {T.~A.}\ \bibnamefont
  {Apostolatos}}, \bibinfo {author} {\bibfnamefont {C.}~\bibnamefont {Cutler}},
  \bibinfo {author} {\bibfnamefont {G.~J.}\ \bibnamefont {Sussman}}, \ and\
  \bibinfo {author} {\bibfnamefont {K.~S.}\ \bibnamefont {Thorne}},\ }\href
  {\doibase 10.1103/PhysRevD.49.6274} {\bibfield  {journal} {\bibinfo
  {journal} {Phys. Rev. D}\ }\textbf {\bibinfo {volume} {49}},\ \bibinfo
  {pages} {6274} (\bibinfo {year} {1994})}\BibitemShut {NoStop}%
\bibitem [{\citenamefont {Hu}\ \emph {et~al.}(2018)\citenamefont {Hu},
  \citenamefont {Li}, \citenamefont {Wang}, \citenamefont {Feng}, \citenamefont
  {Zhou}, \citenamefont {Hu}, \citenamefont {Hu}, \citenamefont {Mei},\ and\
  \citenamefont {Shao}}]{Hu:2018yqb}%
  \BibitemOpen
  \bibfield  {author} {\bibinfo {author} {\bibfnamefont {X.-C.}\ \bibnamefont
  {Hu}}, \bibinfo {author} {\bibfnamefont {X.-H.}\ \bibnamefont {Li}}, \bibinfo
  {author} {\bibfnamefont {Y.}~\bibnamefont {Wang}}, \bibinfo {author}
  {\bibfnamefont {W.-F.}\ \bibnamefont {Feng}}, \bibinfo {author}
  {\bibfnamefont {M.-Y.}\ \bibnamefont {Zhou}}, \bibinfo {author}
  {\bibfnamefont {Y.-M.}\ \bibnamefont {Hu}}, \bibinfo {author} {\bibfnamefont
  {S.-C.}\ \bibnamefont {Hu}}, \bibinfo {author} {\bibfnamefont {J.-W.}\
  \bibnamefont {Mei}}, \ and\ \bibinfo {author} {\bibfnamefont {C.-G.}\
  \bibnamefont {Shao}},\ }\href {\doibase 10.1088/1361-6382/aab52f} {\bibfield
  {journal} {\bibinfo  {journal} {Class. Quant. Grav.}\ }\textbf {\bibinfo
  {volume} {35}},\ \bibinfo {pages} {095008} (\bibinfo {year} {2018})},\
  \Eprint {http://arxiv.org/abs/1803.03368} {arXiv:1803.03368 [gr-qc]}
  \BibitemShut {NoStop}%
\bibitem [{\citenamefont {Cutler}\ and\ \citenamefont
  {Flanagan}(1994)}]{Cutler:1994ys}%
  \BibitemOpen
  \bibfield  {author} {\bibinfo {author} {\bibfnamefont {C.}~\bibnamefont
  {Cutler}}\ and\ \bibinfo {author} {\bibfnamefont {E.~E.}\ \bibnamefont
  {Flanagan}},\ }\href {\doibase 10.1103/PhysRevD.49.2658} {\bibfield
  {journal} {\bibinfo  {journal} {Phys. Rev. D}\ }\textbf {\bibinfo {volume}
  {49}},\ \bibinfo {pages} {2658} (\bibinfo {year} {1994})},\ \Eprint
  {http://arxiv.org/abs/gr-qc/9402014} {arXiv:gr-qc/9402014} \BibitemShut
  {NoStop}%
\bibitem [{\citenamefont {Balasubramanian}\ \emph {et~al.}(1996)\citenamefont
  {Balasubramanian}, \citenamefont {Sathyaprakash},\ and\ \citenamefont
  {Dhurandhar}}]{Balasubramanian:1995bm}%
  \BibitemOpen
  \bibfield  {author} {\bibinfo {author} {\bibfnamefont {R.}~\bibnamefont
  {Balasubramanian}}, \bibinfo {author} {\bibfnamefont {B.~S.}\ \bibnamefont
  {Sathyaprakash}}, \ and\ \bibinfo {author} {\bibfnamefont {S.~V.}\
  \bibnamefont {Dhurandhar}},\ }\href {\doibase 10.1103/PhysRevD.53.3033}
  {\bibfield  {journal} {\bibinfo  {journal} {Phys. Rev. D}\ }\textbf {\bibinfo
  {volume} {53}},\ \bibinfo {pages} {3033} (\bibinfo {year} {1996})},\ \bibinfo
  {note} {[Erratum: Phys.Rev.D 54, 1860 (1996)]},\ \Eprint
  {http://arxiv.org/abs/gr-qc/9508011} {arXiv:gr-qc/9508011} \BibitemShut
  {NoStop}%
\bibitem [{\citenamefont {Chatziioannou}\ \emph {et~al.}(2017)\citenamefont
  {Chatziioannou}, \citenamefont {Klein}, \citenamefont {Yunes},\ and\
  \citenamefont {Cornish}}]{Chatziioannou:2017tdw}%
  \BibitemOpen
  \bibfield  {author} {\bibinfo {author} {\bibfnamefont {K.}~\bibnamefont
  {Chatziioannou}}, \bibinfo {author} {\bibfnamefont {A.}~\bibnamefont
  {Klein}}, \bibinfo {author} {\bibfnamefont {N.}~\bibnamefont {Yunes}}, \ and\
  \bibinfo {author} {\bibfnamefont {N.}~\bibnamefont {Cornish}},\ }\href
  {\doibase 10.1103/PhysRevD.95.104004} {\bibfield  {journal} {\bibinfo
  {journal} {Phys. Rev. D}\ }\textbf {\bibinfo {volume} {95}},\ \bibinfo
  {pages} {104004} (\bibinfo {year} {2017})},\ \Eprint
  {http://arxiv.org/abs/1703.03967} {arXiv:1703.03967 [gr-qc]} \BibitemShut
  {NoStop}%
\bibitem [{\citenamefont {Mangiagli}\ \emph {et~al.}(2019)\citenamefont
  {Mangiagli}, \citenamefont {Klein}, \citenamefont {Sesana}, \citenamefont
  {Barausse},\ and\ \citenamefont {Colpi}}]{Mangiagli:2018kpu}%
  \BibitemOpen
  \bibfield  {author} {\bibinfo {author} {\bibfnamefont {A.}~\bibnamefont
  {Mangiagli}}, \bibinfo {author} {\bibfnamefont {A.}~\bibnamefont {Klein}},
  \bibinfo {author} {\bibfnamefont {A.}~\bibnamefont {Sesana}}, \bibinfo
  {author} {\bibfnamefont {E.}~\bibnamefont {Barausse}}, \ and\ \bibinfo
  {author} {\bibfnamefont {M.}~\bibnamefont {Colpi}},\ }\href {\doibase
  10.1103/PhysRevD.99.064056} {\bibfield  {journal} {\bibinfo  {journal} {Phys.
  Rev. D}\ }\textbf {\bibinfo {volume} {99}},\ \bibinfo {pages} {064056}
  (\bibinfo {year} {2019})},\ \Eprint {http://arxiv.org/abs/1811.01805}
  {arXiv:1811.01805 [gr-qc]} \BibitemShut {NoStop}%
\bibitem [{\citenamefont {Baird}\ \emph {et~al.}(2013)\citenamefont {Baird},
  \citenamefont {Fairhurst}, \citenamefont {Hannam},\ and\ \citenamefont
  {Murphy}}]{Baird:2012cu}%
  \BibitemOpen
  \bibfield  {author} {\bibinfo {author} {\bibfnamefont {E.}~\bibnamefont
  {Baird}}, \bibinfo {author} {\bibfnamefont {S.}~\bibnamefont {Fairhurst}},
  \bibinfo {author} {\bibfnamefont {M.}~\bibnamefont {Hannam}}, \ and\ \bibinfo
  {author} {\bibfnamefont {P.}~\bibnamefont {Murphy}},\ }\href {\doibase
  10.1103/PhysRevD.87.024035} {\bibfield  {journal} {\bibinfo  {journal} {Phys.
  Rev. D}\ }\textbf {\bibinfo {volume} {87}},\ \bibinfo {pages} {024035}
  (\bibinfo {year} {2013})},\ \Eprint {http://arxiv.org/abs/1211.0546}
  {arXiv:1211.0546 [gr-qc]} \BibitemShut {NoStop}%
\end{thebibliography}%
\end{document}